\documentclass{article}

\usepackage{PRIMEarxiv}
\usepackage{authblk} 
\usepackage[utf8]{inputenc} 
\usepackage[T1]{fontenc}    
\usepackage{hyperref}       
\usepackage{url}            
\usepackage{booktabs}       
\usepackage{amsfonts}       
\usepackage{nicefrac}       
\usepackage{microtype}      
\usepackage{lipsum}
\usepackage{float}
\usepackage{multirow}%
\usepackage{graphicx}
\usepackage{siunitx}
\usepackage{lineno}
\usepackage{booktabs}
\usepackage{flafter} 
\usepackage{tcolorbox}
\usepackage{xspace}
\usepackage{adjustbox}
\usepackage{standalone}
\usepackage{tikz}
\usepackage{tabularx}
\usepackage{placeins}
\usepackage{amssymb}
\usepackage{amsmath}
\usepackage{subcaption}
\usepackage{esdiff}
\usepackage[title]{appendix}%
\usepackage[backend=biber, style=numeric, sorting=none]{biblatex}

\usepackage{algorithm}
\usepackage{algpseudocode}   

\graphicspath{{Images/}}     

\makeatletter
\newcommand{\resetAuthors}{%
  \renewcommand\AB@authlist{}%
  \renewcommand\AB@affillist{}%
  \renewcommand\AB@authors{}%
  \renewcommand\AB@affilsep{\protect\Affilfont}%
  \renewcommand\AB@blk@and{\protect\Authfont\protect\AB@setsep}%
}
\makeatother

\title{Particle Identification Studies using a Highly Granular Crystal Electromagnetic Calorimeter}

\author[a,b]{J.\,Rolph\thanks{corresponding author},}
\author[a,b]{Y.\,Liu,}
\author[a,b]{B.\,Qi\thanks{formerly at IHEP/UCAS}}

\affil[a]{Institute of High Energy Physics, Chinese Academy of Sciences (IHEP, CAS),\\
  19B Yuquan Road, Shijingshan District, Beijing 100049, China}

\affil[b]{University of Chinese Academy of Sciences (UCAS),\\
  19A Yuquan Road, Shijingshan District, Beijing 100049, China}

\author[d,c,e]{S.\,Li,}
\author[d,e]{Z.\,Zhao,}
\author[c,d,e]{H.\,Yang}

\affil[c]{School of Physics and Astronomy, Shanghai Jiao Tong University,\\
  800 Dongchuan Road, Shanghai 200240, China}

\affil[d]{Tsung-Dao Lee Institute, Shanghai Jiao Tong University,\\
  1 Lisuo Road, Shanghai 201210, China}

\affil[e]{National Key Laboratory of Dark Matter Physics; Shanghai Key Laboratory for Particle Physics and Cosmology;\\
  Key Laboratory for Particle Astrophysics and Cosmology (MoE), Shanghai 200240, China}

\bibliography{main}

\begin{document}
\maketitle


\begin{abstract}
A proposal for the Circular Electron Positron Collider electromagnetic calorimeter is a highly granular scintillating crystal design that uses silicon photomultipliers to measure physics signals from photons. The potential to develop particle identification methods using deep and traditional machine learning is explored. A contrastive machine learning technique was found to enable the separation of showering and 'punch-through' pions without explicit labels. A supervised particle identification method was then trained using the results, using both a graph neural network and a boosted decision tree to distinguish electron, showering hadron, and minimum-ionizing-particle-like events. The classifier achieved an area-under-the-receiver operating characteristic curve of 0.996 for both cases. A Shapley additive explanations analysis showed the boosted decision tree classifier matched expectations for each event type. The particle identification capability is then demonstrated on experimental data taken from the CERN Proton Synchrotron T9 beamline.
\end{abstract}


\section{Introduction}

Future Higgs factories such as the Circular Electron-Positron Collider (CEPC) \cite{group_cepc_2025} and Future Circular Electron-Positron Collider (FCC-ee) \cite{benedikt_future_2025}  will precisely measure the properties of the Higgs boson, weak vector bosons, and the top quark, while also performing critical validations of the Standard Model. A typical figure of merit for the required performance of Particle Flow is to reconstruct the mass of the Higgs boson with \qtyrange{3}{4}{\percent} resolution. This approach necessitates accurate tracking of charged particles within a jet, sophisticated event reconstruction techniques, and highly granular sampling calorimeters. Furthermore, the electromagnetic calorimeter (ECAL) must possess a resolution sufficient to reconstruct the energy of photons, while also being designed with enough granularity to resolve individual particles with excellent energy and time resolution. 

Particle identification (PID) is possible using the highly granular energy density information. This information can be used to enhance event reconstruction. Examples of relevant physics measurements that rely on PID are jet tagging (e.g., Higgs reconstruction from hadronic decays (e.g. $H\rightarrow q\bar{q}\/\tau\bar{\tau}$), from photon decays (e.g., $H\rightarrow \gamma \gamma)$ and $\pi^{0}$ reconstruction ($B^{0}\/B^{0}_{s} \rightarrow \pi^{0}\pi^{0}$, relevant for CP-violation studies \cite{wang_prospects_2022}). Jet energy resolution can also be improved through superior charged particle and track association by Particle Flow Algorithms (PFAs). 

Hadrons may interact with the calorimeter either by ionizing and then showering, or by ionizing the medium only. These events are called showering hadrons and 'punch-through' hadrons, respectively. Punch-through hadrons behave generally indistinguishably from muons or other MIPs at the momenta expected for future Higgs factories. A standard method of identifying MIP-like events is to build an ad hoc selection criterion for a MIP-like event by assuming a cut criterion. However, this method is limited as the final classification will only be as accurate as the initial definition and may introduce bias. For instance, challenging examples to define in a cut-based analysis could be when the ionizing track through the calorimeter is split between cells or when there is partial showering by either hadrons or muons. This motivates research into novel ways of defining subsets from data with at least two classes (e.g., showering and MIP-like pion), for which a pure sample of one class is available (muon/antimuon). 

Two studies are performed. First, an unsupervised contrastive deep machine learning method is used to distinguish MIP-like pions from showering pions. Then, using the learned representation of MIP-like and showering pions, particle identification (PID) models are developed using deep machine learning and boosted decision tree classifiers (BDT), which are then compared.

\section{Methods and Tools}

\subsection{Highly-Granular Crystal ECAL Prototype}

The High-Granularity Crystal Calorimeter (HGCCAL) is the primary choice for the CEPC ECAL system \cite{group_cepc_2025}. A prototype has been developed to test the performance. Each layer of the crystal ECAL prototype consists of 14 scintillating crystal bars per layer, each of \qtyproduct[product-units = single]{12x12x2}{\centi \meter \cubed} volume. Each subsequent layer is rotated orthogonally with respect to the previous one. This ’cross-hatch’ structure enables reconstruction of the energy density of deposits in the crystal by combining information from two orthogonal and adjacent layers and transverse granularity while minimizing dead volume and the number of readout channels compared to a design with single crystals. Scintillation light from the crystal is read out from each end by two SiPMs. There are 14 layers of the calorimeter. The first 12 layers are composed of beryllium germanium oxide (BGO). The last 2 layers are composed of bismuth silicate (BSO), which acts as a tail-catcher. The total depth of the calorimeter is \qty{24.9}{X_{0}}.

\begin{figure}
    \centering
    \includegraphics[width=\linewidth]{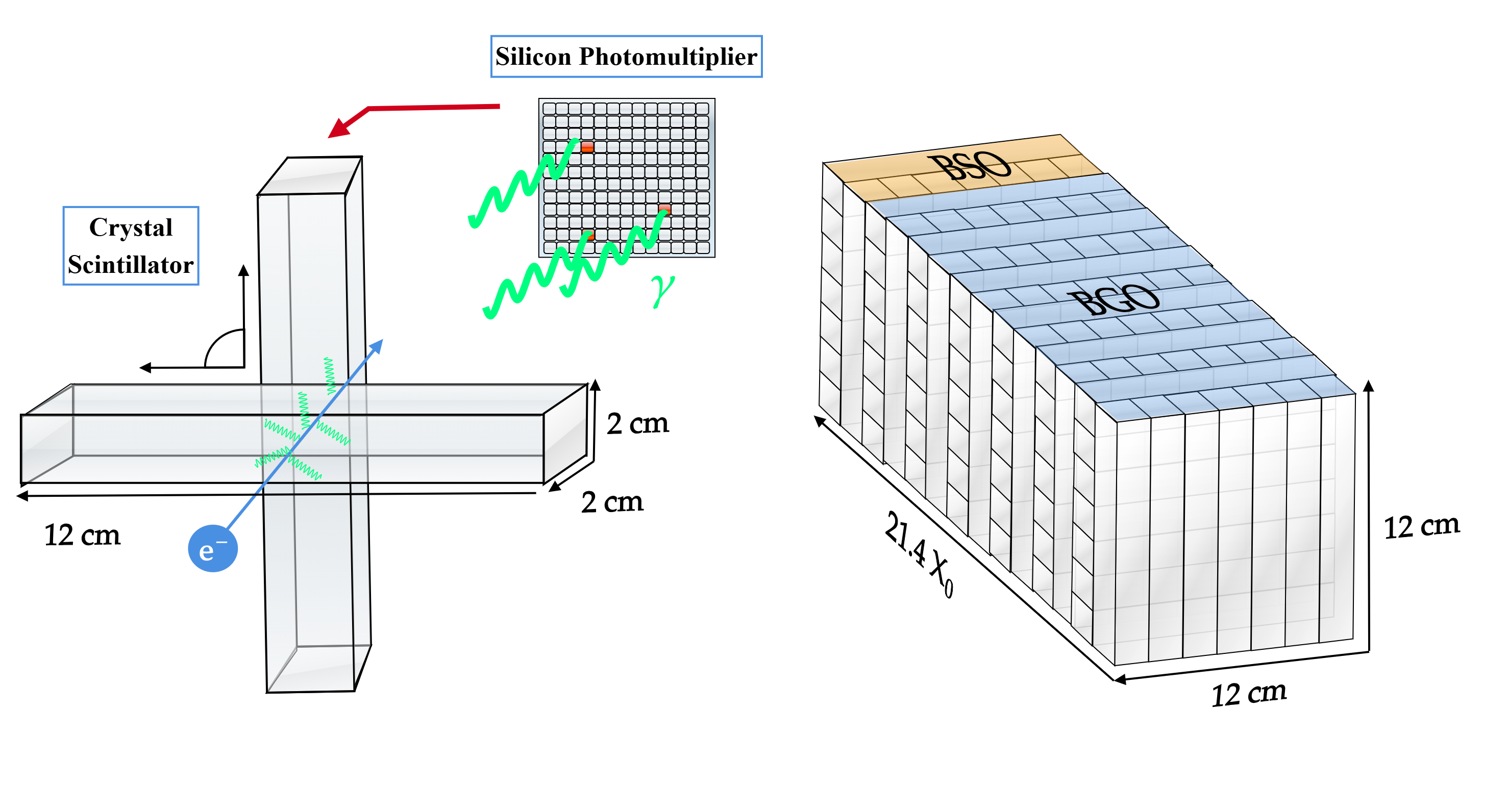}
    \caption{Diagram illustrating the design of the HGCCAL calorimeter.}
    \label{fig:fig_hgccal_diagram}
\end{figure}

During 2025, the calorimeter was used in a test beam campaign at the Proton Synchrotron (PS) and Super Proton Synchrotron (SPS) at CERN in Geneva, Switzerland. The entire system was configured in a low-energy beamline setup with all upstream beam instrumentation removed to minimize particle scattering and momentum spread. A total of 20.8 million  events were collected across an energy range of \qtyrange{0.1}{10}{\giga \electronvolt}. However, low beam purity for certain energies meant that samples of electrons were heavily contaminated by (anti)muons and hadrons in the range \qtyrange{6}{10}{\giga \electronvolt}. This motivated the development of a PID algorithm in order to purify the data.

\begin{figure}
    \centering
    \includegraphics[width=0.8\linewidth]{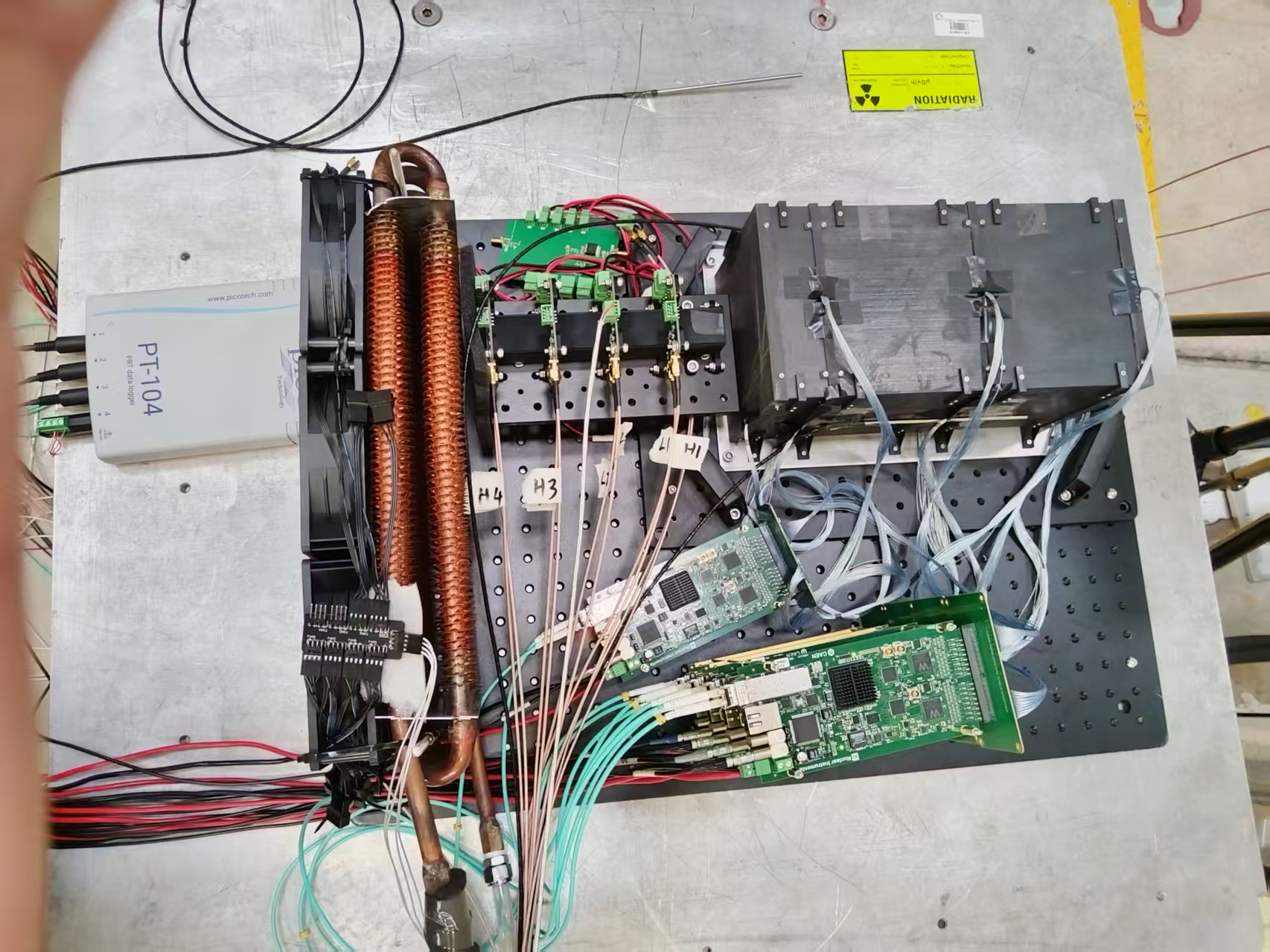}
    \caption{Photograph of the HGCCAL at the test beam at PS in 2025.}
    \label{fig:fig_hgcal_tb}
\end{figure}

A photograph of the calorimeter at the test beam is shown in Fig.~\ref{fig:fig_hgcal_tb}. 

\subsubsection{Datasets}

\paragraph{Simulation}

\begin{figure}
\centering
\includegraphics[width=0.8\linewidth]{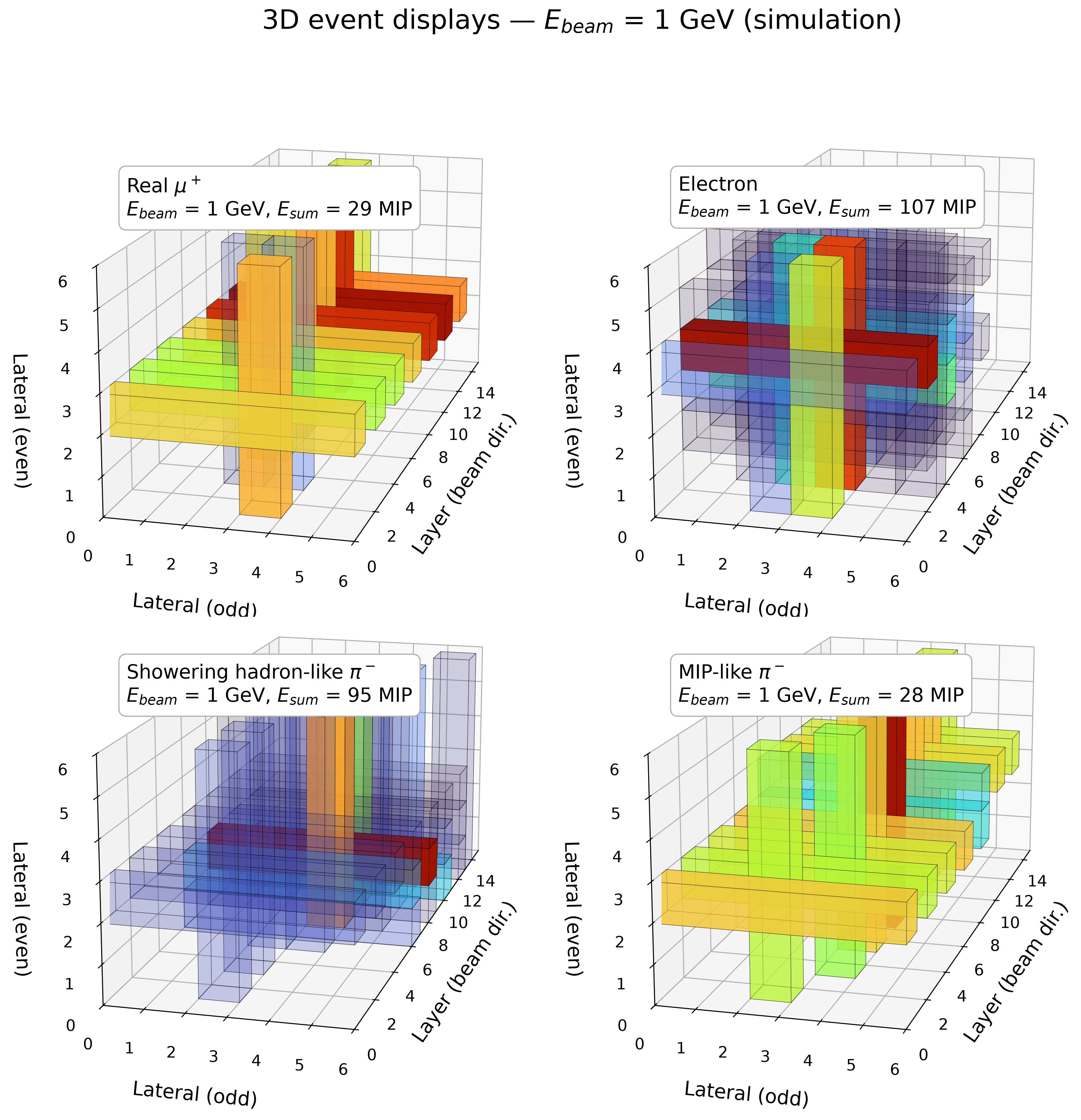}
\caption{Event displays of interacting particles from the simulation dataset. The crystal bars are indicated by the cuboidal bars. The color (blue to red) and opacity (clear to opaque) indicate higher vs. lower energy deposits in the bar. The $x$ and $y$ axes indicate the odd and even layers, and the $z$-axis indicates the depth. The top left and right plots show an electron and anti-muon event, and the bottom left and right events show a showering pion and punch-through pion, respectively. The energy sums are shown in the legend.}
\label{fig:fig_event_displays_3d}
\end{figure}

A \texttt{Geant4} simulation \cite{agostinelli_geant4simulation_2003} of electron, pion, and anti-muon events was used in the study, in the energy range from \qtyrange{0.1}{10}{\giga \electronvolt}. The raw simulation was digitized to model experimental and detector effects. Firstly, the digitization accounts for scintillation effects: the Poisson fluctuations of the number of scintillation photons, the attenuation of the light reaching the SiPMs due to self-absorption in the crystal, and the photon detection efficiency (PDE) of the sensors. The effective light yield has been determined as \qty{200}{\mathrm{p.e.} \per \mathrm{MIP}} ($\qty{200}{\mathrm{p.e}} / \qty{8.9}{\mega \electronvolt}$). Secondly, SiPM effects such as saturation and cross-talk are considered. Lastly, the effect of the readout electronics is considered, which affects the conversion of the SiPM charge to analog-to-digital (ADC) channels. Three gain levels are used to obtain a broad dynamic range of the sensors. The threshold between low and medium/high gain is \qty{8e3}{\mathrm{p.e}}. The signal is also convolved with electronic noise as a normal distribution centered at \qty{0}{\mathrm{p.e.}} with an experimentally obtained scale of \qty{1.56}{\mathrm{p.e.}}. The simulation was then calibrated to units of \unit{\giga \electronvolt}. A table of statistics is available in Appendix Table~\ref{tab:event_counts_sim}. Example events from the simulation are visualized in Fig.~\ref{fig:fig_event_displays_3d}.

\paragraph{Experimental Data}

During 2025, experimental data were taken at the CERN PS-T9 beamline using the crystal ECAL prototype. During this test beam, upstream instrumentation, such as SiFi trackers and Čerenkov detectors, was removed to minimize the beam momentum spread and improve resolution measurements. Energies from \qtyrange{5}{10}{\giga \electronvolt} of the campaign are studied henceforth and are used only to demonstrate the performance of the PID applied to data, as these were the most heavily contaminated of the samples available. The \qty{5}{\giga \electronvolt} sample was expected to be a relatively pure sample of electrons. By contrast, in the range \qtyrange{6}{10}{\giga \electronvolt}, considerable contamination by (anti)muons and hadrons was expected.  

The data were reconstructed using a standard analysis procedure for calorimeters with SiPM-based readout. The SiPM response is measured in 'analog-to-digital' counts (ADC counts). Two ASIC operation modes were measured simultaneously during the test beam: high-gain and low-gain ADC values. Measurements of the pedestal, $\mathrm{ADC}_\mathrm{ped}$  and MIP-scale $\mathrm{ADC}_\mathrm{MPV}$, are required for energy reconstruction. The pedestal was measured for each readout channel using a double-sided Crystal Ball distribution (DSCB) fit of the SiPM pedestal  for both high- and low-gain modes. The DSCB is introduced in Eq.~\ref{eq:DSCB} in Appendix Sec.~\ref{sec:non_standard_distributions}. The mode of the fit yields the pedestal value, $\mathrm{ADC}_{\mathrm{ped}}$. A normal distribution was also found to be consistent. The MIP-scale was measured using $\mu^{+}$ events. Ionizing energy deposits were selected by rejecting pedestal energy deposits based on the fitted distribution of the pedestal, $\mathcal{L}(\operatorname{DSCB}(\mathrm{ADC} | \theta_{\mathrm{ped}})) < -10$ and $\mathrm{ADC} > \mathrm{ADC}_{\mathrm{ped}}$, where $\theta_{\mathrm{ped}}$ are the parameters from the pedestal fit. All subsequent layers were required to have a single energy  $\pm 1$ bar index (lateral/horizontal), to ensure a well-defined, straight track per event. A Landau-Gaussian convolution was fitted to each channel to measure the MPV for the high-gain mode, $\mathrm{ADC}_{\mathrm{MPV, HG}}$.

 The high- and low-gain modes are non-linear and must be corrected. Events are selected from a high-purity \qty{5}{\giga \electronvolt} dataset containing predominantly $e^{-}$ events,  and the joint distribution of the high- and low-gain ADC values is examined. Firstly, a locally weighted scatterplot smoothing (LOWESS) spline \cite{cleveland_smoothing_1996, josef_perktold_statsmodelsstatsmodels_2024} was fitted to the joint distribution to extract the overall profile of the non-linearity. Then, a reference linear relationship between high- and low-gain modes was reconstructed by selecting the spline region with the least curvature and requiring the line to pass through the pedestals of both modes. The low-gain MPV value, $\mathrm{ADC}_{\mathrm{MPV, LG}}$, is then estimated from the line based on the value of $\mathrm{ADC}_{\mathrm{MPV, HG}}$. Finally, the correction for the low-gain mode is obtained using a translation with a spline, $C_{\mathrm{corr}}(\mathrm{ADC}_\mathrm{HG},  i_{\mathrm{channel}})$, which maps the spline to the reference line. An additive constant is used, as it does not affect the variance of the low-gain ADC value around the new line.

 The energy is then reconstructed according to Eq. ~\ref{eq:energy_reco}
 
\begin{equation}
E_{\mathrm{MIP}} =
\begin{cases}
    \dfrac{\mathrm{ADC}_{\mathrm{HG}} - \mathrm{ADC}_{\mathrm{Ped, HG}}}{\mathrm{ADC}_{\mathrm{MPV, HG}}} 
        & \text{if } \mathrm{ADC}_{\mathrm{HG}} < 7900~\text{counts} \\[1em]
    \dfrac{\mathrm{ADC}_{\mathrm{LG}} + C_{\mathrm{corr}}(\mathrm{ADC}_{\mathrm{HG}}, i_{\mathrm{channel}}) - \mathrm{ADC}_{\mathrm{Ped, LG}}}{\mathrm{ADC}_{\mathrm{MPV, LG}}} 
        & \text{otherwise}
\end{cases}
\label{eq:energy_reco}
\end{equation}

 A table of statistics is available in Appendix Table~\ref{tab:event_counts_data}.

\subsection{Classifier Frameworks}

\subsubsection{GravNet}

\texttt{GravNet} is a graph neural network architecture introduced in Ref.~\cite{qasim_learning_2019} for learning on point clouds. It was explicitly designed for Particle Flow event reconstruction in highly granular calorimeters and is therefore used in this study. Point clouds are defined as points representing some information in Cartesian space. In the context of calorimeters, these 'features' are the energy deposited or its timestamp, while the position of the sensor measuring the energy defines the graph. This model was chosen as it has demonstrated superior performance compared to applicable alternatives such as \texttt{DGCNN} and \texttt{PointNet} in Particle Flow reconstruction in highly granular calorimeters. 

\begin{figure}
    \centering
    \includegraphics[width=\linewidth]{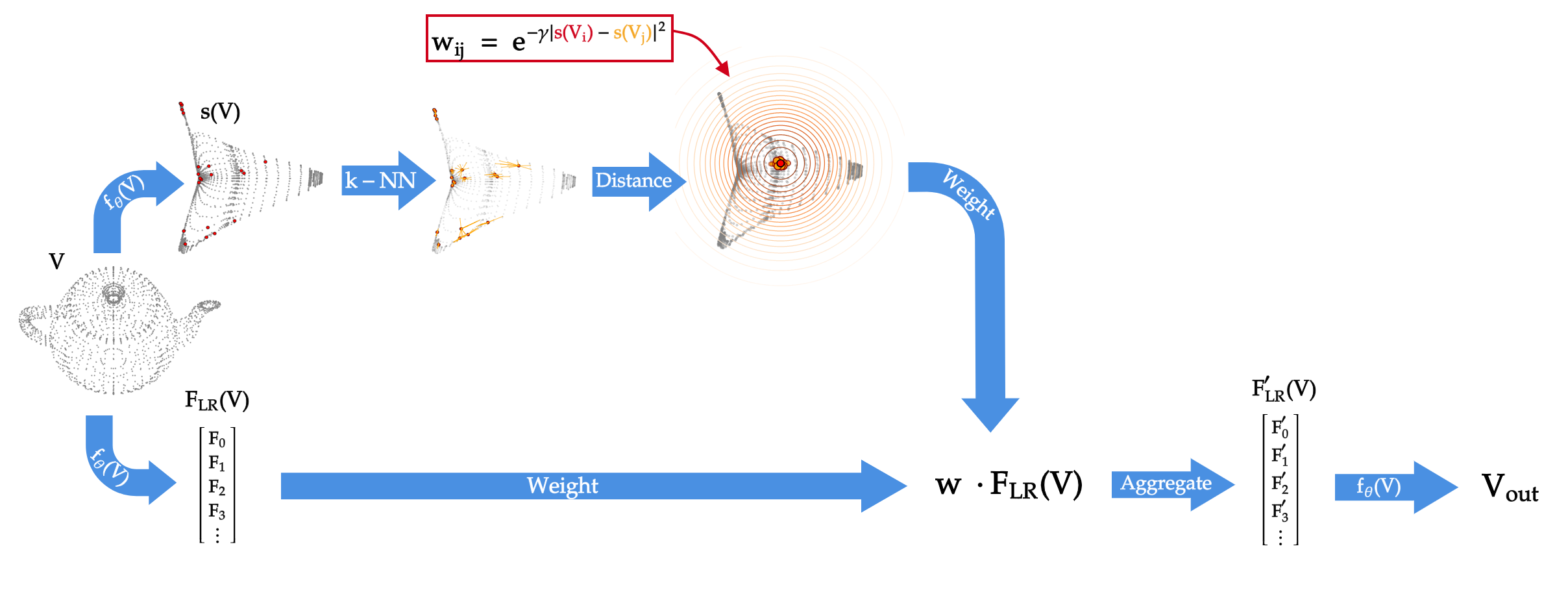}
    \caption{Flow chart diagram illustrating the architecture of \texttt{GravNet}. $V$ are some input vertices indicated by the Utah teapot. $f(V; \theta)$ indicates a convolutional layer with learnable parameters $\theta$. $k$-NN indicates the $k$ nearest neighbours.}
    \label{fig:fig_gravnet}
\end{figure}

\texttt{GravNet} operates by clustering using the $k$-nearest neighbors of each point in a projected subspace, $s$, where proximity in the space represents similarity in the context defined by the optimization. The Gaussian function is then used to calculate weights for each neighbour cluster, $w_{ij} = \exp \left( -\gamma \cdot d(s_i, s_j)^{2} \right)$, $d(u, v)$ is the L2-norm between two vectors $u$ and $v$, $d(u, v) = \lVert u - v \rVert_2 $, and $s_i$, $s_j$   An additional learned feature space $F_{\mathrm{LR}}$ is obtained, representing learned properties about each point in the data. $F_{\mathrm{LR}}$ is then weighted according to all the weights obtained for each cluster (i.e., each point is scaled by $k$ weights from the subspace) and aggregated over the cluster dimension. In this study, the maximum, the mean, and the second moment, representing unnormalized variance, are used as aggregation functions. A final output layer projects the output to a new representation of the graph. The concept is illustrated in Fig.~\ref{fig:fig_gravnet} 

\texttt{GravNet} operations are typically applied recursively to obtain deeper representations of the underlying distributions of the data. 
The \texttt {PyTorch Geometric} \cite{fey_fast_2019} and \texttt {PyTorch} \cite{paszke_pytorch_2019} software are used to implement this framework in the study.

\subsubsection{Boosted Decision Trees}

Boosted decision trees (BDTs) are a commonly used machine learning model for classification. They utilize an ensemble of many weak decision trees. A decision tree is a learning model that splits data using recursively applied, learnable thresholds and is typically shallow. Ensembles are used because they are less sensitive to outliers than a single strong classifier. The trees are aggregated to produce a prediction.

'Gradient boosting' refers to optimizing using the gradient of the loss with respect to the ensemble's predictions. This is achieved by fitting a new tree using the gradients and adding it to the model during training. It has been demonstrated that this is a more accurate learning method and generalizes better than a single decision tree \cite{friedman_greedy_2001}. They are often applied in particle physics as classifier models as they are typically interpretable and robust.

The \texttt{xGBoost} framework \cite{chen_xgboost_2016} is used to implement boosted decision trees in the study.

\subsection{Methods}

\subsubsection{Unsupervised PID for Label Extraction for Pions}
\label{sec:UnsupervisedPIDforLabelExtractionforPions}

 Hadrons interacting with a calorimeter are generally expected to either shower or act as minimum  ionizing particles and punch through. The set of events recorded by the calorimeter can be defined as $\pi \in M \cup \overline{M}$, where $M$ is the set of MIP-like hadrons and $\overline{M}$ is the set of showering hadrons or other types of events, such as partially showering hadrons or other particles. By contrast, muons are drawn nearly always from $M$, with a very small probability of showering.

A discriminative representation of $M$ can be learned by requiring two criteria. Firstly, any two samples drawn from $M$ should be 'close' in a chosen metric. Secondly, $M \cup \overline{M}$ should be 'far' from $M$ in that metric. Since punch-through hadron events, represented in the study by MIP-like pions in simulation, are drawn from the subset of $M$ within $M \cup \overline{M}$ and are indistinguishable and constrained to remain close to the muon events drawn from $M$, optimizing both conditions necessarily pushes elements of $\overline{M}$, which are showering pions in simulation, further from $M$. In summary, if a pure set of one class can be obtained, then it can be extracted from a mixed set without prior knowledge of the difference. This concept is referred to as contrastive learning.

A method known as a Siamese neural network is typically used for contrastive learning. This is a configuration of a machine learning model that takes multiple inputs during training with the same weights and outputs a latent vector for each input. The network is optimized to draw examples of the same class closer and push other classes further apart by comparing the outputs of the neural network between similar and dissimilar samples of inputs.

The loss function is called 'triplet loss' \cite{balntas_learning_2016}. An 'anchor' vector, $a \in M$, is used as a reference point for comparisons. Then, two further samples, a positive and a negative sample ($p \in M$, $n \in M \cup \bar{M}$ ), are selected. The loss is then defined according to Eq.~\ref{eq:triplet_loss}:

\begin{equation}
    \mathcal{L}(a,p,n)=\max\{d(a,p)-d(a,n)+\text{margin},0\},
    \label{eq:triplet_loss}
\end{equation}

where $\mathcal{L}$ is the loss, $d(u, v)$ is the L2-norm between two vectors $u$ and $v$, $d(u, v) = \lVert u - v \rVert_2 $. The maximum is taken over one mini-batch used for training. The fundamental idea is that vectors of the positive class are drawn towards the anchor class, while those from the negative class are pushed away from it. The maximum function selects the most challenging case in the batch to speed up convergence and is called 'mining'. For this reason, the anchor and positive samples were randomly sampled from the $M$ dataset and the negative from the $\bar{M}$ dataset for every triplet.

It is noted that the neural network may learn a trivial solution to the contrastive problem by increasing the magnitude of the latent vector to become arbitrarily large. For instance, the network may enter a failed convergence state where the vectors of one class are much longer than the other, which means the loss does not improve in a meaningful way. The vector is normalised using the L2-norm to avoid this problem, which constrains the distance to $d(u, v) = 2\sqrt{1 - \cos{\theta}}$, $d(u,v) \in \left[0, 2\right)$. 

Inference is achieved by finding the latent vectors from the training dataset of the target class and taking a normalized average. This average vector is called the 'prototype'. The cosine similarity between the positive class and the prototype then represents the intra-class variation. This can be used to determine a cutoff criterion to classify events as either positive or negative in the mixed dataset.

\begin{figure}
    \centering
    \includegraphics[width=0.5\linewidth]{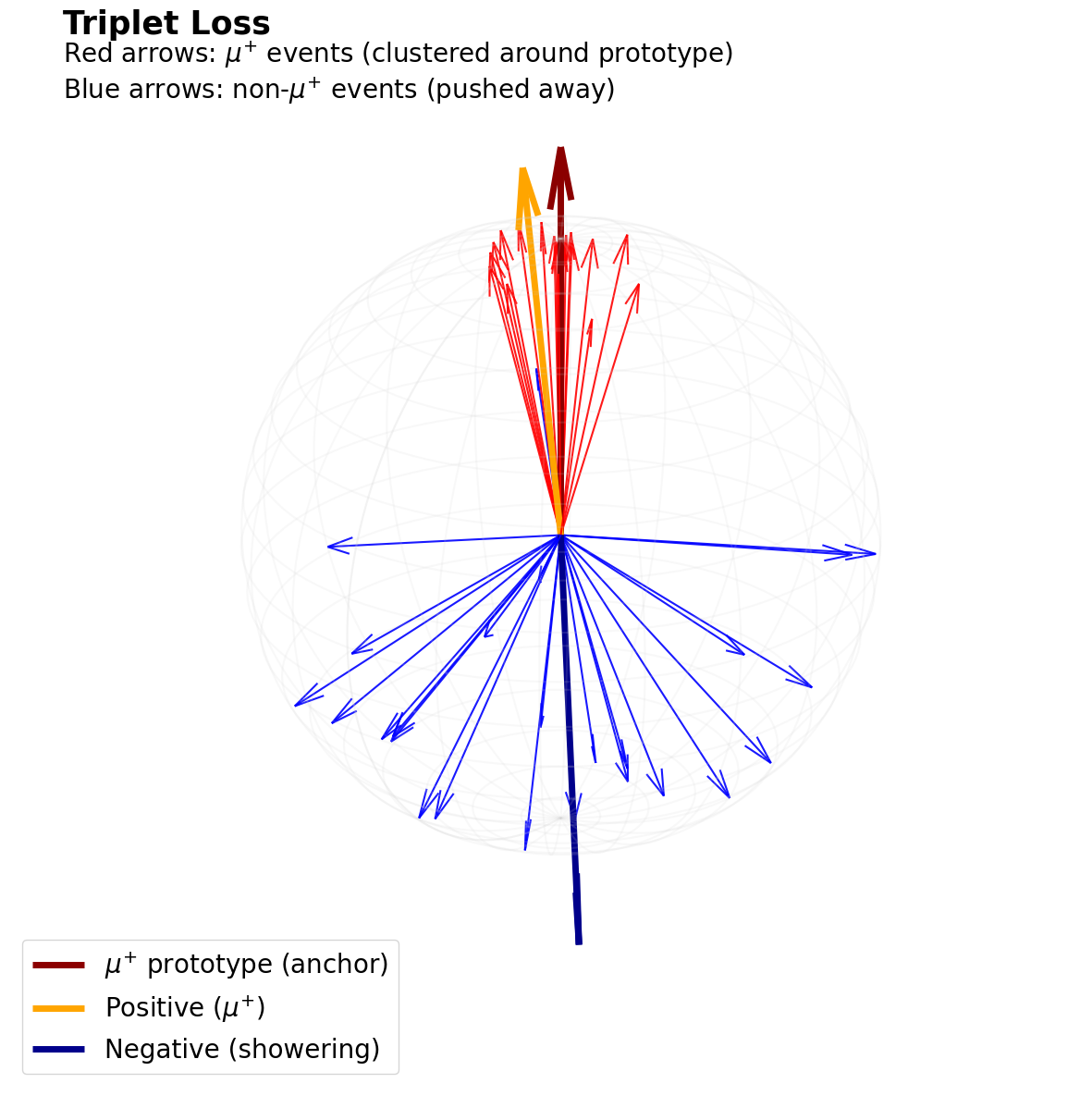}
    \caption{Diagram illustrating the concept of contrastive triplet loss training. Red vectors indicate the positive (target) class ($M$) and the blue vectors represent the negative class ($\bar{M}$). The deep red line indicates the anchor. The red lines indicate other $\mu$ events. The blue lines indicate showering hadrons.  }
    \label{fig:contrastive}
\end{figure}

\subsubsection{Supervised PID for Performance Studies}

The previous section provides a method for obtaining labels for the MIP-like pion sample. After this is achieved, the goal of the study is to assess the particle identification performance of the ECAL prototype. The model should classify three groups of events: 'electron', 'MIP-like', and 'showering hadron-like'. The showering hadron fraction is determined using the results of the Siamese neural network method presented in Sec.~\ref{sec:UnsupervisedPIDforLabelExtractionforPions}, the results of which are discussed in Sec.~\ref{sec:UnsupervisedPIDResults}.

PID was performed using both the \texttt{GravNet} and BDT models, using cross-entropy loss: 

\begin{equation}
    \mathcal{L}(y, p) = -\frac{1}{N}\sum_{i=1}^{N}\sum_{j=1}^{C} y_{ij}\log(p_{ij})
\end{equation}

where $\mathcal{L}$ is the loss, $N$ is the number of events (either over a mini-batch for the neural network or the entire dataset for the BDT), $C$ is the number of classes (3), with indices $i$ and $j$, respectively. $y_{ij}$ is the ground-truth label ($y_{ij}=1$ if event $i$ belongs to class $j$) and $p_{ij}$ is the predicted probability of that class by the model. Each model predicts logits ($z$), and the probability is reconstructed using the Softmax operator $\sigma(z)$:

\begin{equation}
    \sigma(z_{j}) = \frac{\exp(z_j)}{\sum_j \exp(z_j)}
\end{equation}

\subsection{Training and Validation}

\subsubsection{Unsupervised PID}\label{sec:SiameseNeuralNetwork}

A Siamese neural network consisting of five successive \texttt{GravNet} layers, using 8 spatial dimensions for the clustering space and 32 dimensions for the feature space. 16 neighbors are used for clustering. The input to the neural network is the natural logarithm of one plus the energy deposited in a bar, which is performed to constrain the highly skewed energy distribution while remaining in the domain $E_{\mathrm{hit}}\in \left[0, \infty \right)$, and the graph nodes are defined by the crystal bar index and layer. All intermediate layer outputs are concatenated to preserve multi-scale features, which are then aggregated via mean and second-moment pooling across the event. These pooled features pass through a two-layer multi-layer perceptron (MLP) with 512 and 256 hidden units, incorporating \qty{20}{\percent} dropout for regularization. The outputs are L2-normalized 128-dimensional embeddings where same-class particles cluster and different classes separate in the latent space. In all cases, the leaky ReLU activation with a default negative-value gradient of 0.01 is used.

Two models were trained: one to distinguish $e^{-}$ from $\mu^{+}$ as a control method, and another to distinguish $\pi^{-}$ from $\mu^{+}$. A validation sample of 1000 events of each particle type at every particle energy was used. The model learning rate was optimized using the method of Ref.~\cite{smith_cyclical_2017}, using the \texttt{torch\_lr\_finder} software library \cite{noauthor_torch-lr-finder_nodate}. The ADAM optimizer \cite{kingma_adam_2017} and an exponential learning rate decay were used to optimize.

After training converged, a global prototype, $p$, for both models was determined by taking the normalized mean of a random subset of \num{1e5} embedding vectors from the training dataset on the target class, $\mu^{+}$. The validation set containing each particle was then projected onto the prototype, $\cos{\theta} = u \cdot p$. This maps the vectors to a cosine similarity in the domain $\left[-1,1\right)$. A cutoff criterion was determined from the intra-class distribution of $\mu^{+}$, using the right-tailed confidence interval $P(\cos{\theta} \geq X) = \qty{99}{\percent}$, performed for each classifier separately. 

Firstly, the discrimination power of the model was tested using the $e^{-}$ vs. $\mu^{+}$ model. This trivial separation task illustrates that the model functions as intended. The confusion matrix and Receiver Operating Characteristic (ROC) curve show the true positive rate as a function of the false positive rate of the classifier:

\begin{subequations}
    \begin{equation}
         \mathrm{TPR} = \frac{\mathrm{TP}}{\mathrm{TP} + \mathrm{FN}}
    \end{equation}
    
    \begin{equation}
         \mathrm{FPR} = \frac{\mathrm{FP}}{\mathrm{FP} + \mathrm{TN}}
    \end{equation}
\end{subequations}

Where $\mathrm{TPR}$ and $\mathrm{FPR}$ are the true and false positive rates, $\mathrm{TP}$ and $\mathrm{FP}$ are the numbers of true and false positives, and $\mathrm{TN}$ and $\mathrm{FN}$ are the numbers of true and false negatives. The area under the curve (AUC) measures performance, $\text{AUC} \in \left[0.5, 1\right)$, where 0.5 is a random classifier, and 1.0 is perfect.

The same study was then performed for the $\pi^{-}$ vs. $\mu^{+}$ model, with the expectation that some of the $\pi^{-}$ events cannot be distinguished from the $\mu^{+}$ events. These 'misidentified' events are then compared to MIP-like $\mu^{+}$ events. As a ground-truth label is not available in this case, agreement between MIP-like $\pi^{-}$ events and $\mu^{+}$ events is measured in several ways. Fits of the Landau-Gaussian distribution to the energy deposited per layer are used to measure the agreement with the expectation for ionizing energy loss. If the cut effectively selects MIP-like events, the agreement of the Landau-Gaussian distributions with the layer-wise energy distributions should be good; the most probable value for the energy loss per layer should be equal across layers and approximately the same for the 'misidentified' $\pi^{-}$ events and the $\mu^{+}$ sample. It was noted that below around \qty{0.5}{\giga \electronvolt}, the $\mu^{+}$ sample did not achieve full penetration of the calorimeter. The \qty{0.4}{\giga \electronvolt} sample is nonetheless presented to illustrate this effect. To assess the showering part of the sample, shower shape fits were used to compare the measured value of the pion nuclear interaction length for BGO with the literature. Finally, the fraction of events predicted to 'punch through' the calorimeter can be used to assess the agreement with expectations for the MIP-like behavior of the $\pi^{-}$ sample:
\begin{equation}
    P_{\overline{S}} = e^{-\frac{L}{\lambda_{\pi}}}
    \label{eq:punch_through_probability}
\end{equation}

Where $L$ is the longitudinal depth of the calorimeter, $L=\qty{28}{\centi \meter}$, and $\lambda_{\pi}$ is the pion interaction length of BGO, $\lambda_{\pi} = \qty{26.72}{\centi \meter}$ \cite{pdg_review_2026, pdg_review_2026_url}.

\subsubsection{Supervised PID}\label{sec:NeuralNetworkandBoostedDecisionTreeClassifiers}

 The classifiers are designed to separate 3 classes of events: 'electron', 'showering hadron ', and 'MIP-like' events. From the raw dataset, $e^{-}$ and $\mu^{+}$ are labeled 'electron' and 'MIP-like' by default. The $\pi^{-}$ particles are labeled using the Siamese network developed in Sec.~\ref{sec:SiameseNeuralNetwork}. The \qty{99}{\percent} right-tailed percentile of the prototype projection distribution is used as a cut criterion. Events with a projected value below the criterion are labeled as 'showering hadron-like', and above the criterion they are labeled as 'MIP-like '. Justification for this is given in the results presented in Sec.~\ref{sec:UnsupervisedPIDResults}. A separate validation sample of \num{1000} events for every event class and available particle energy was selected, totaling \num{84000} events. 

Two classifier models were trained to test the performance and validity of different classifier methods. The first was a \texttt{GravNet} network with the same structure as described in Sec.~\ref{sec:SiameseNeuralNetwork}, with separate weights, used as the classifier model. The only difference is in the final layer, which produces 3 logits for each event class. All training and hyperparameter optimization details were otherwise the same as for that network.  As the computational resources required for hyperparameter tuning are considerably less for a BDT than a deep neural network, the hyperparameters (learning rate, maximum tree depth, fraction of the training data to randomly sample without replacement before growing each tree, and fraction of variables to randomly select for each tree) were tuned using the Optuna hyperparameter tuning library \cite{akiba_optuna_2019}. The second was a standard BDT that uses a simple set of standard global variables for showers as input, shown in Table~\ref{tab:bdt_params}. The distributions of each parameter are shown for the \qty{0.2}{\giga \electronvolt}, \qty{1}{\giga \electronvolt} and \qty{5}{\giga \electronvolt} samples in Appendix Fig.~\ref{fig:fig_BDT_features_by_class_E0p2},  Appendix Fig.~\ref{fig:fig_BDT_features_by_class_E1} and Appendix Fig.~\ref{fig:fig_BDT_features_by_class_E5}, respectively. The moments of  the central log ratio (CLR) of the bar energies in a shower used in the classifier are the standard deviation ($\sigma_{\mathrm{clr}}$), skewness ($\gamma_{\mathrm{clr}}$), and excess kurtosis ($\kappa_{\mathrm{clr}}$). The CLR is defined as:

\begin{equation}
    \operatorname{CLR}(E_{\mathrm{bar}}) = \log(E_{\mathrm{bar}}) - \langle \log(E_{\mathrm{bar}}) \rangle, \qquad E_{\mathrm{bar}} > 0
\end{equation}

CLR is chosen for representing shower fluctuations, as the transformation is invariant under additive shifts in log space and is therefore independent of the total energy deposited in an event. This has two key advantages: no energy scale is required for the classifier, which means that the model is never exposed to the limits of the energy spectrum; it provides good agreement when comparing simulation to data, for which MIP-energy scaling may be slightly different.

\begin{table}[]
    \centering
    \renewcommand{\arraystretch}{1.2}
    \setlength{\tabcolsep}{12pt}
    \caption{Event-level aggregated parameters used as inputs to the BDT. 
    The table lists the variable name, symbol, and definition for each. 
    Here, $i$ is the bar index (left$\to$right, up$\to$down) running from 1 to 6, 
    $l$ is the layer index from 1 to 14, $E_{\mathrm{sum}}$ is the total deposited energy in the event, 
    $N_{\mathrm{hits}}$ is the number of bars above threshold, and ``even''/``odd'' refer to even/odd bar indices (for transverse) or layers (for longitudinal). $f_{\text{layer}, l-1}$ is defined such that $f_{\text{layer}, 0}$ corresponds to the first layer. $m_{i}$ refers to the $i^{\text{th}}$ statistical moment. }
    \begin{tabular}{c|cc}
        \toprule
        Variable Name                  & Symbol        & Definition \\
        \midrule
        CLR Variance            & $\sigma_{\mathrm{clr}}$     &  $\sqrt{m_{2}(\operatorname{CLR}(E_{\mathrm{bar}}))}$\\
        CLR Skewness            & $\gamma_{\mathrm{clr}}$     &  $\frac{m_{3}(\operatorname{CLR}(E_{\mathrm{bar}})}{m_{2}(\operatorname{CLR}(E_{\mathrm{bar}})^{1.5}}$\\
        CLR Excess Kurtosis            & $\kappa_{\mathrm{clr}}$     &  $\frac{m_{4}(\operatorname{CLR}(E_{\mathrm{bar}})}{m_{2}(\operatorname{CLR}(E_{\mathrm{bar}})^{2}}$\\
        
        Center of Gravity, $x$    & $\mathrm{CoG}_{X}$ & $\dfrac{\sum_{i=2, \text{even}} E_{\mathrm{bar},i} \cdot i}{E_{\mathrm{sum, even}}}$ \\
        Center of Gravity, $y$         & $\mathrm{CoG}_{Y}$ & $\dfrac{\sum_{i=1, \text{odd}} E_{\mathrm{bar},i} \cdot i}{E_{\mathrm{sum, odd}}}$ \\
        Center of Gravity, $z$         & $\mathrm{CoG}_{Z}$ & $\dfrac{\sum_{l=1} E_{\mathrm{bar},l} \cdot l}{E_{\mathrm{sum}}}$ \\
        Mean Distance from Center of Gravity, $x$ & $\langle |X - \mathrm{CoG}_{X}| \rangle$ & $\frac{\sum_{i=2, \text{even}} | i - \text{CoG}_{x} |} {N_{\mathrm{hit, even}}}$ \\
        Mean Distance from Center of Gravity, $y$ & $\langle |Y - \mathrm{CoG}_{Y}| \rangle$ 
        & $\frac{\sum_{i=1, \text{odd}} | i - \text{CoG}_{y} |}{N_{\mathrm{hit, odd}}}$ \\
        Fraction of Shower Energy In Layer $l$  & $f_{\text{layer}, l-1}$ &  $\frac{E_{\mathrm{bar},l}}{E_{\mathrm{sum}}}$ \\
                                     
        \bottomrule
    \end{tabular}
 
    \label{tab:bdt_params}
\end{table}

\section{Results}

\subsection{Unsupervised PID}\label{sec:UnsupervisedPIDResults}

Firstly, the results of the $e^{-}$ vs. $\mu^{+}$ model are presented to demonstrate the classification capacity of the model using ground truth labels. Then, the results of the $e^{-}$ vs. $\pi^{-}$ model are presented and analysed.

\subsubsection{$e^{-}$ vs. $\mu^{+}$}

\begin{figure}
\centering

\begin{subfigure}{0.49\linewidth}
\centering
\includegraphics[width=\linewidth]{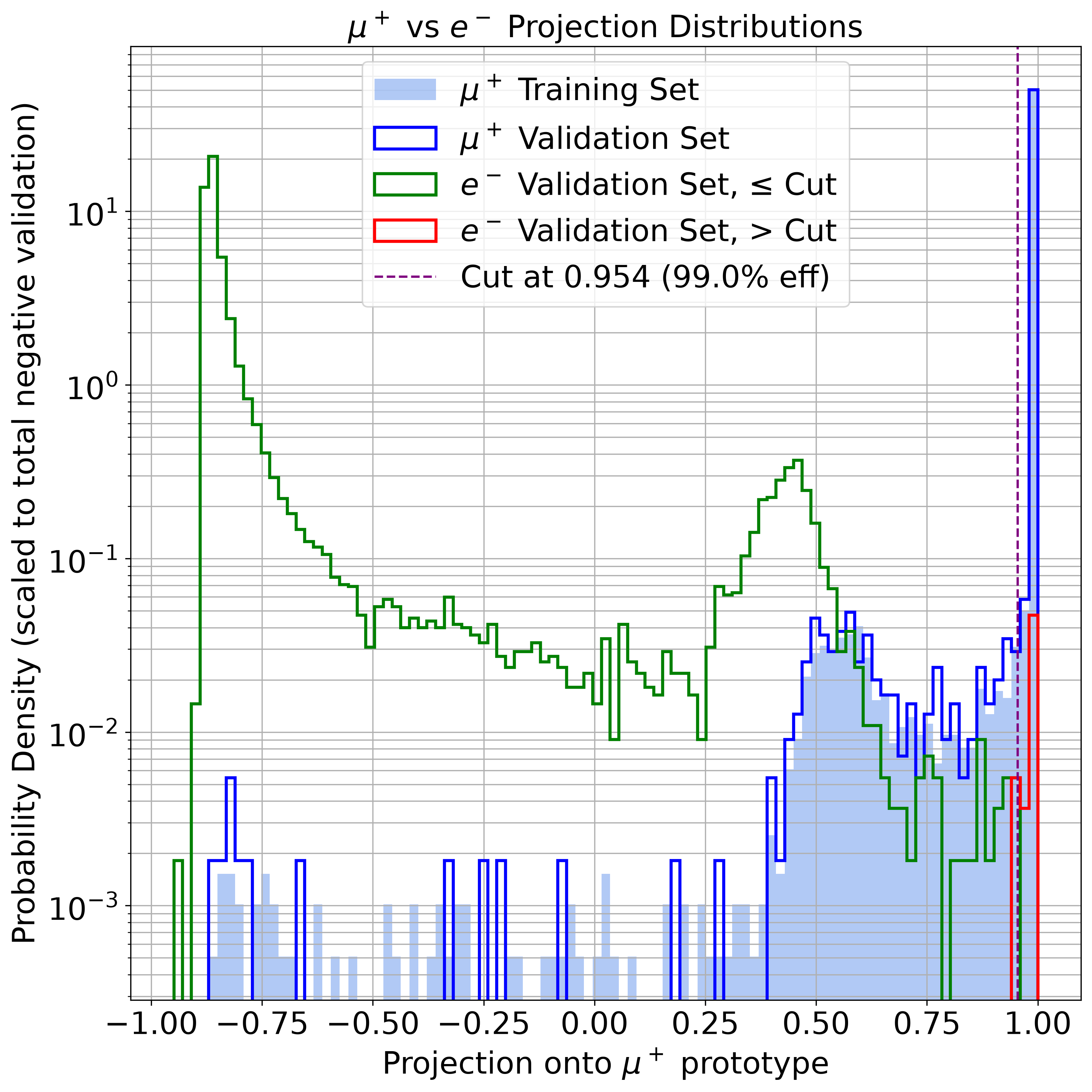}
\caption{}
\label{fig:fig_proj_e-_vs_mu+}
\end{subfigure}

\vfill

\begin{subfigure}{0.49\linewidth}
\centering
\includegraphics[width=\linewidth]{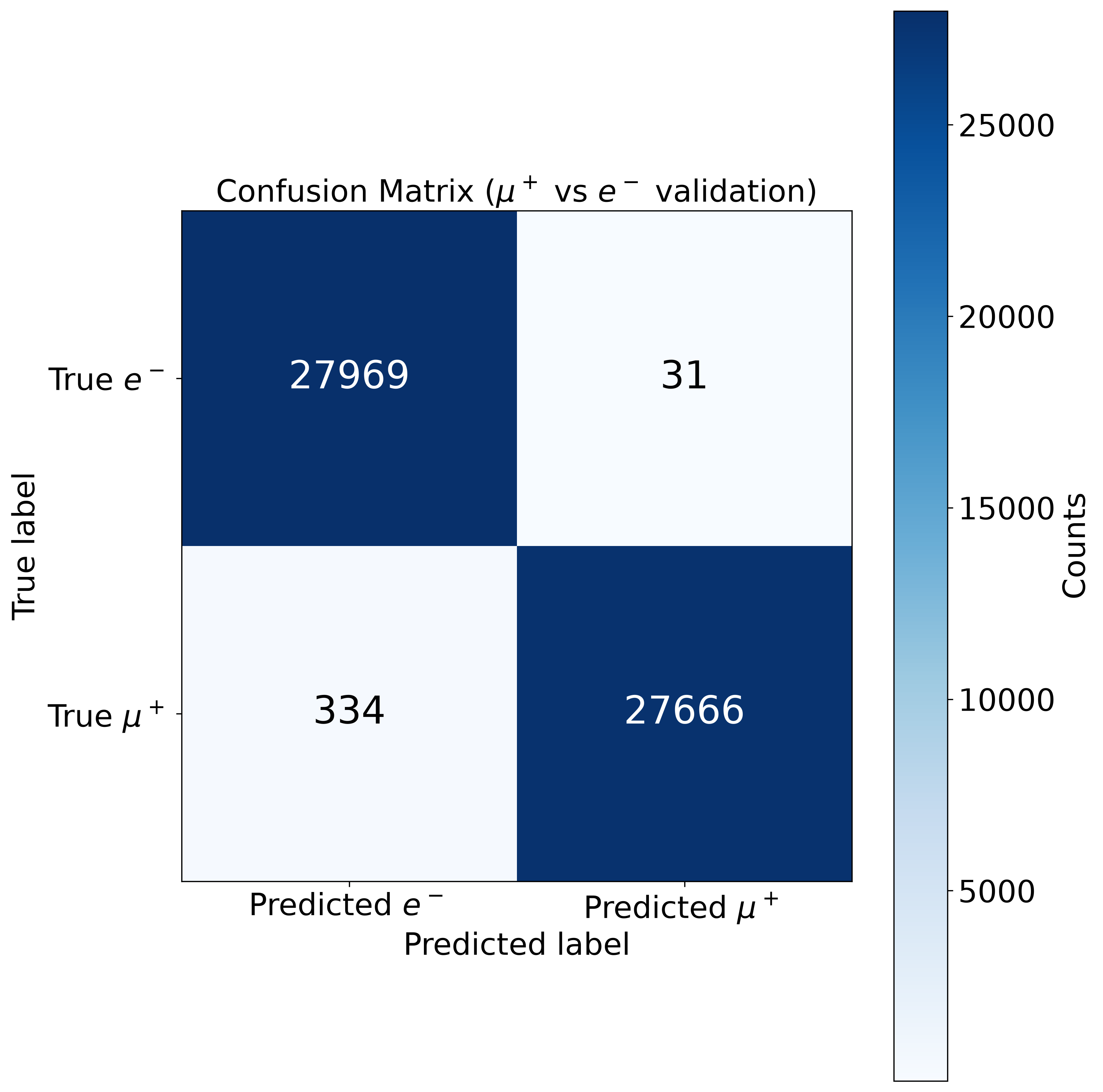}
\caption{}
\label{fig:fig_confusion_matrix_e-_vs_mu+}
\end{subfigure}
\hfill
\begin{subfigure}{0.49\linewidth}
\centering
\includegraphics[width=\linewidth]{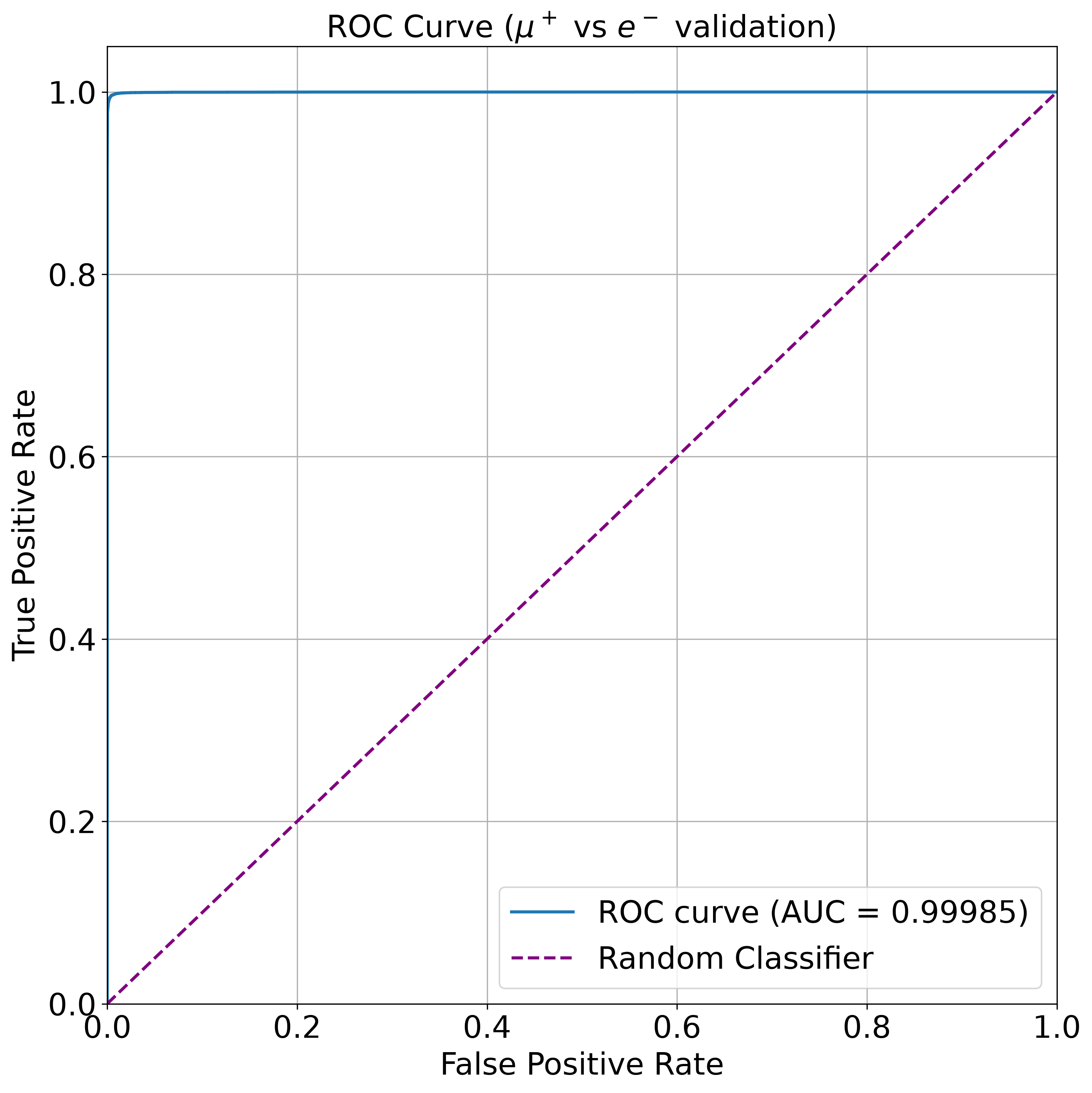}
\caption{}
\label{fig:fig_roc_e-_vs_mu+}
\end{subfigure}

\caption{Performance metrics of the unsupervised $e^{-}$ vs. $\mu^{+}$ model. Fig.~\ref{fig:fig_proj_e-_vs_mu+} shows the distribution of the projections onto the prototype for the positive training and validation datasets ($\mu^{+}$), shown in light shaded blue and blue. The vertical purple line indicates the \qty{99}{\percent} right-tailed percentile ('cut') of the positive training dataset (shaded blue region). The negative validation sample less than or equal to the cut value (showering hadron) is shown in green, and the negative validation sample greater than the cut value (MIP-like) is shown in red. Fig.~\ref{fig:fig_confusion_matrix_e-_vs_mu+} shows the confusion matrix of the classifier using the particle labels. White to blue indicates fewer to more events. Fig.~\ref{fig:fig_roc_e-_vs_mu+} shows the ROC curve of the classifier using the particle labels. The blue line shows the ROC curve, and the purple dashed line shows the expectation for a random classifier. The AUC is shown in the legend.}
\label{fig:fig_e-_vs_mu+}
\end{figure}

The distribution of the projected value onto the prototype for the training and validation positive class ($\mu^{+}$), and the validation negative class ($e^{-}$), the selected cut value, confusion matrix, and ROC curve are presented for the $\mu^{-}$ vs. $e^{-}$ model in Fig.~\ref{fig:fig_e-_vs_mu+}. 

Fig.~\ref{fig:fig_proj_e-_vs_mu+} shows the distribution of the $\mu^{+}$ and $\pi^{-}$ projections on the prototype in the hyperspherical latent space of the Siamese network. A mostly bimodal distribution is observed with clear peaks at $\cos{\theta} \simeq -1$ (not MIP-like) and $\cos{\theta} \simeq 1$ (MIP-like). This indicates the two classes are positioned at antipodal poles in the latent space, with each pole corresponding to one of the two classes. A small third peak is observed for the $\pi^{-}$ sample at around  $\cos{\theta} \simeq 0.5$. This third peak is observed most strongly for the \qty{0.1}{\giga \electronvolt} sample and is practically negligible at higher energies. The peak corresponds to electron events with roughly the same number of active bars in the event as a $\mu^{+}$ event of the same energy, and indicates that these 'low activity' events are closer to MIP-like events than normal showering electrons. However, it is noted that this third peak does not considerably overlap with the $\mu^{+}$ projection distribution, and therefore does not considerably affect the effectiveness of the discrimination cut at $\cos{\theta} = 0.954$.

Fig.~\ref{fig:fig_confusion_matrix_e-_vs_mu+} and Fig.~\ref{fig:fig_roc_e-_vs_mu+} shows the confusion matrix and ROC for $e^{-}$ and $\mu^{+}$ classification, respectively. The muon classifier has a muon purity of \qty{98.1}{\percent} and rejects \qty{99.81}{\percent} of electrons. The AUC for the classifier is greater than \qty{99.9}{\percent}. This provides clear evidence that the model works effectively as intended as a model for a basic particle classification task.

\subsubsection{$\pi^{-}$ vs. $\mu^{+}$}

\begin{figure}
\centering

\begin{subfigure}{0.49\linewidth}
\centering
\includegraphics[width=\linewidth]{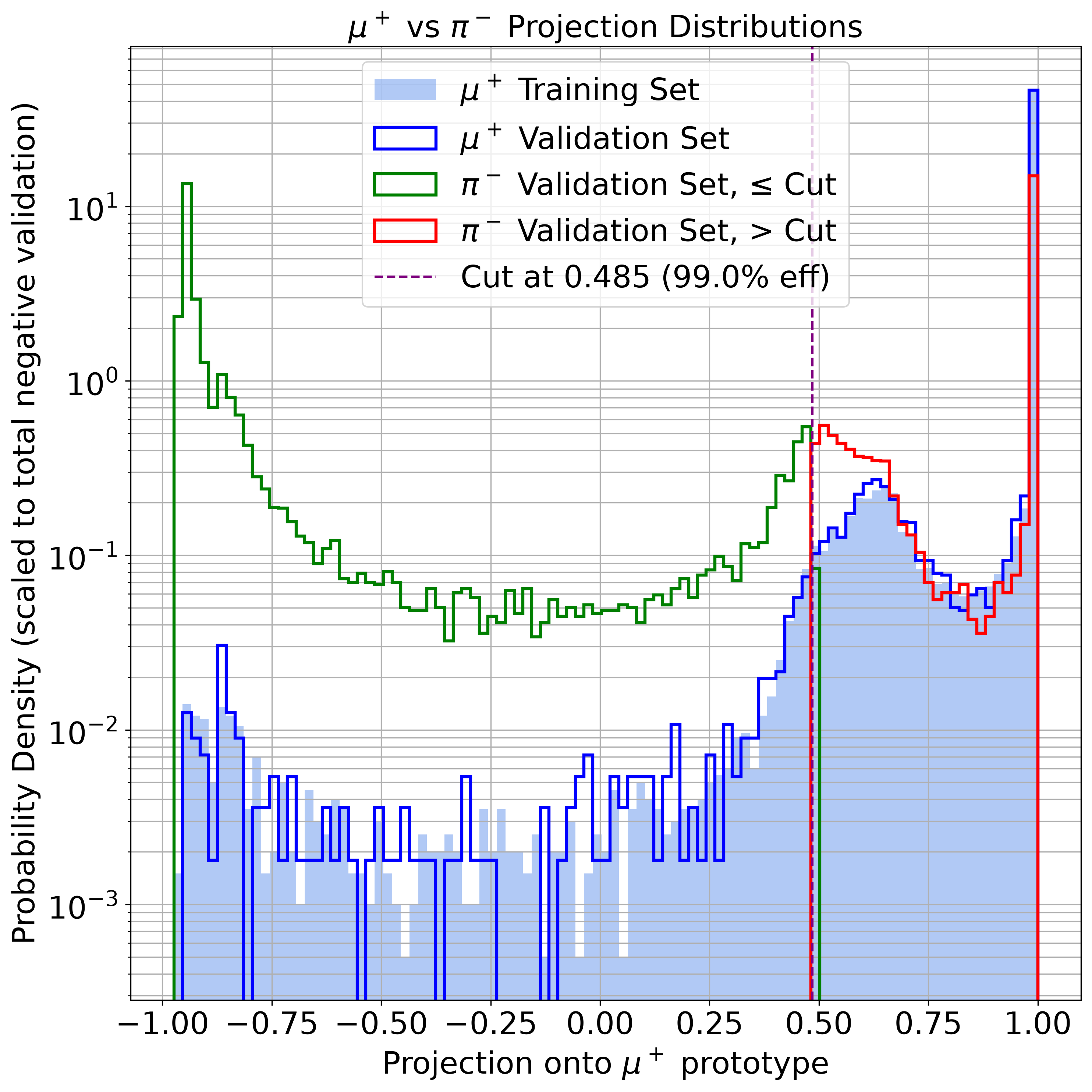}
\caption{}
\label{fig:fig_proj_pi-_vs_mu+}
\end{subfigure}

\vfill

\begin{subfigure}{0.49\linewidth}
\centering
\includegraphics[width=\linewidth]{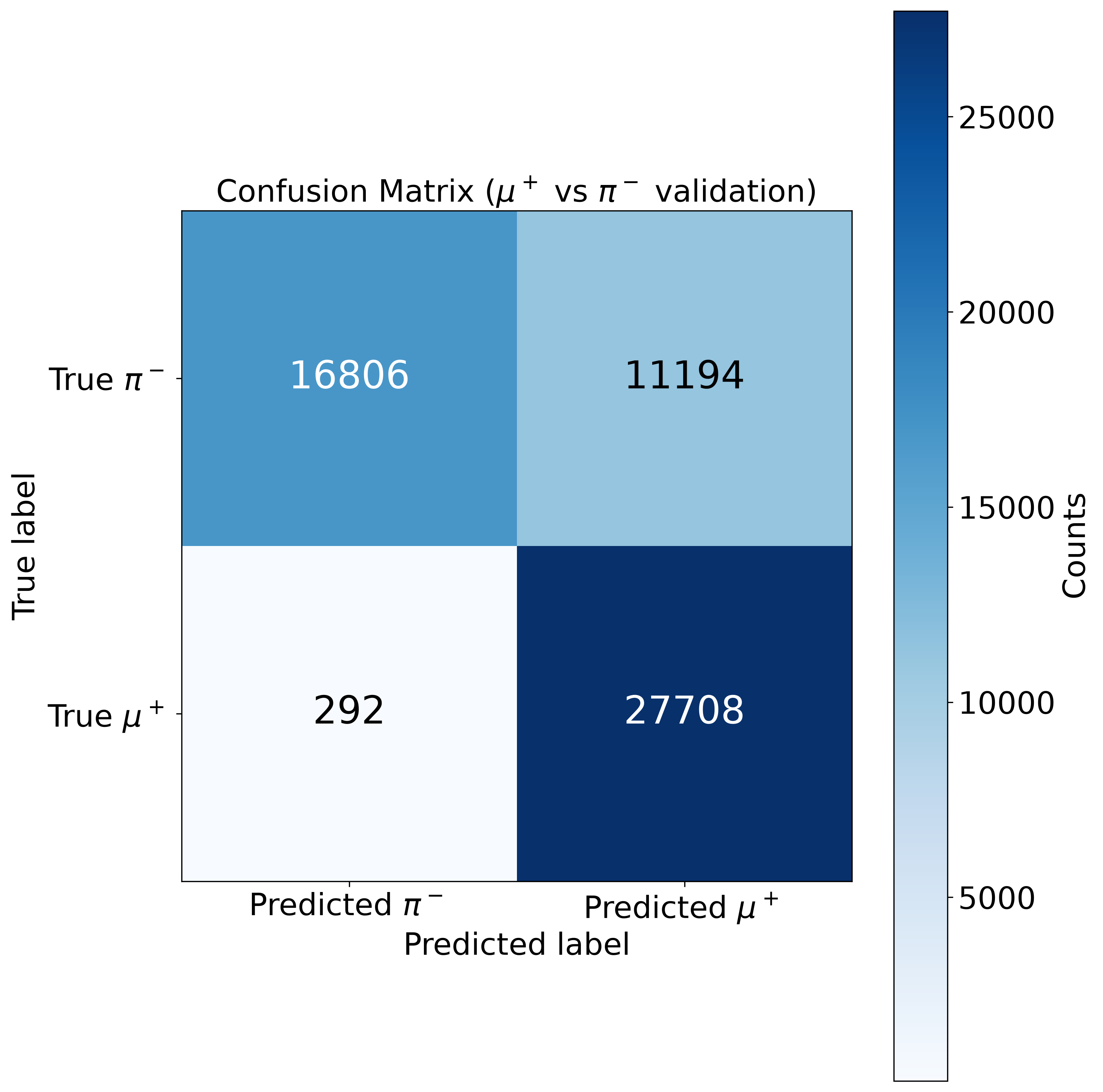}
\caption{}
\label{fig:fig_confusion_matrix_pi-_vs_mu+}
\end{subfigure}
\hfill
\begin{subfigure}{0.49\linewidth}
\centering
\includegraphics[width=\linewidth]{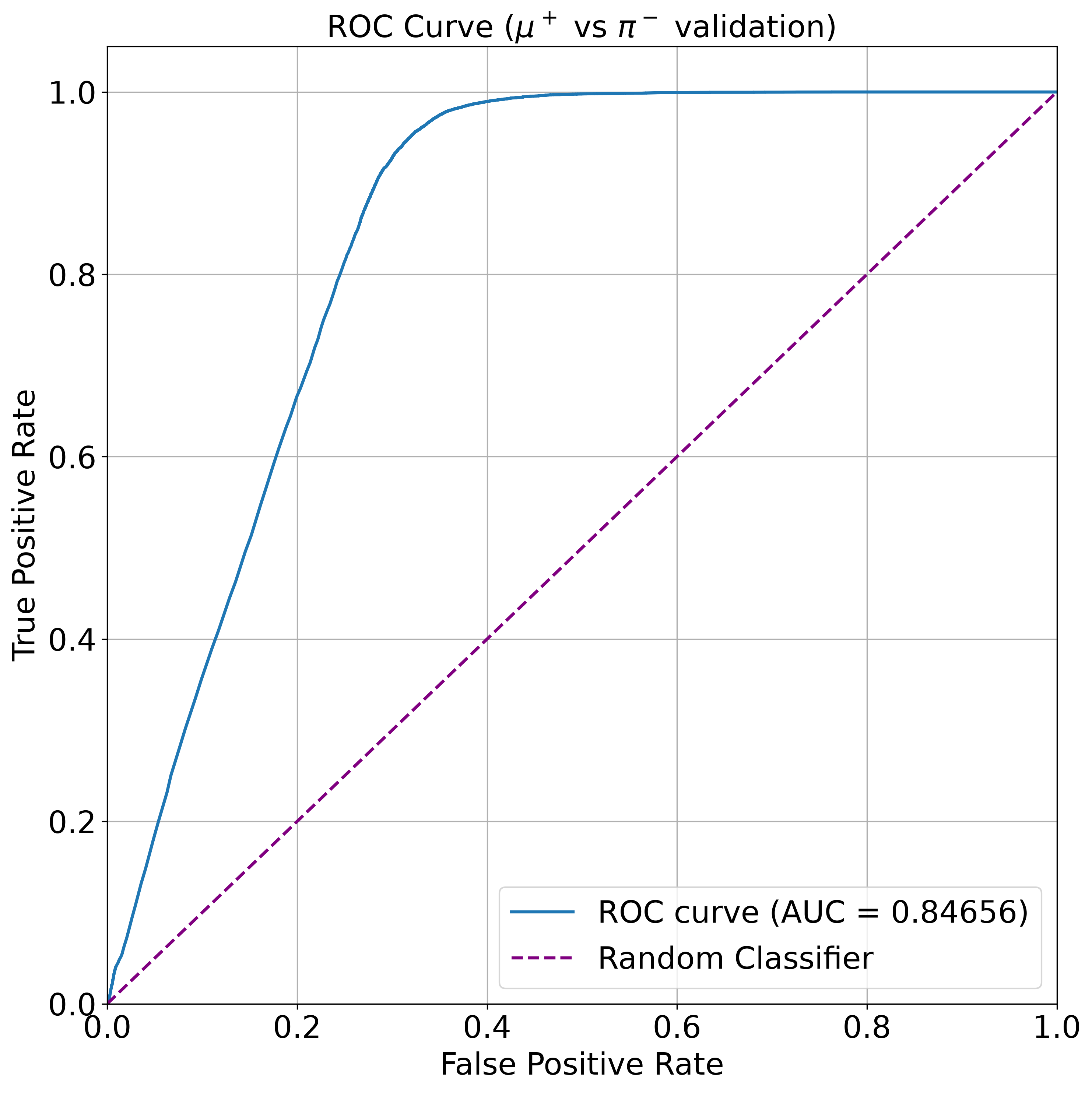}
\caption{}
\label{fig:fig_roc_pi-_vs_mu+}
\end{subfigure}

\caption{Performance metrics of the unsupervised $\pi^{-}$ vs. $\mu^{+}$ model. Otherwise, figures correspond to descriptions of Fig.~\ref{fig:fig_e-_vs_mu+}. }
\label{fig:fig_pi-_vs_mu+}
\end{figure}

The same results are first presented for the $\pi^{-}$ vs. $\mu^{+}$ in Fig.~\ref{fig:fig_pi-_vs_mu+}.
as for the $e^{-}$ vs. $\mu^{+}$ in Fig.~\ref{fig:fig_e-_vs_mu+}. Fig.~\ref{fig:fig_proj_pi-_vs_mu+} shows a similar result as for the $e^{-}$ vs. $\mu^{+}$ case in Fig.~\ref{fig:fig_proj_e-_vs_mu+}, indicating antipodal poles for each class. However, a considerable fraction of the total sample of events matches the distribution of $\mu^{+}$, indicated by the red histogram where $\cos{\theta} > 0.485$. Again, a third peak is observed at around $\cos{\theta} \simeq 0.5$, and was again found to correspond to 'low activity' $\pi^{-}$ events with similar numbers of active bars to the $\mu^{+}$ sample. 

Comparison with the confusion matrix in Fig.~\ref{fig:fig_roc_pi-_vs_mu+} reveals a $\mu^{+}$ efficiency of \qty{98.6}{\percent} but a purity of only \qty{71.2}{\percent}. The large fraction indicates irreducible background, which can be interpreted as MIP-like hadrons, of which a total of \qty{40}{\percent} are $\pi^{+}$ passing the cut across all particle energies. This latter fraction can be interpreted as punch-through pions, which are presently studied.

Energy deposited by ionization is expected to follow the Landau-Gaussian distribution, defined according to Eq. ~\ref{eq:landau_gaus}
\begin{equation}
\operatorname{LandauGaus}(x; \mu, \sigma, \text{MPV}, \xi) = \int_{-\infty}^{\infty} L(x^{\prime}; \mathrm{MPV}, \xi) \cdot \mathcal{N}(x - x^{\prime}; \mu=0, \sigma) dx^{\prime}
\label{eq:landau_gaus}
\end{equation}

where $L$ and $\mathcal{N}$ are the Landau and Gaussian distributions, $\mu$ and $\sigma$ are the Gaussian mean and width parameters, and $\text{MPV}$ and $\xi$ are the most probable value and scale parameter of the Landau distributions. The Landau distribution is calculated via approximation according to Ref.~\cite{yoshimura_numerical_2024}.  The convoluted density function is determined numerically.

The overall distribution of deposited energy in each bar and the corresponding distribution per layer were found for each of the 'real' $\mu^{+}$ events and the $\pi^{-}$ events passing or failing the cut, defined as 'MIP-like' or 'showering hadron-like' $\pi^{-}$ events, respectively. Events were selected with particle energies greater than \qty{1}{\giga \electronvolt}, below which MIP-like events were observed to range out in the calorimeter due to reaching the Bragg region of the Bethe-Bloch equation. The distributions were then fitted using the Landau-Gaussian distribution with the \texttt{MIGRAD} algorithm of \texttt{iminuit} using extended binned log-likelihood \cite{dembinski_scikit-hepiminuit_2025}.  

\begin{figure}
    \centering
    \includegraphics[width=\linewidth]{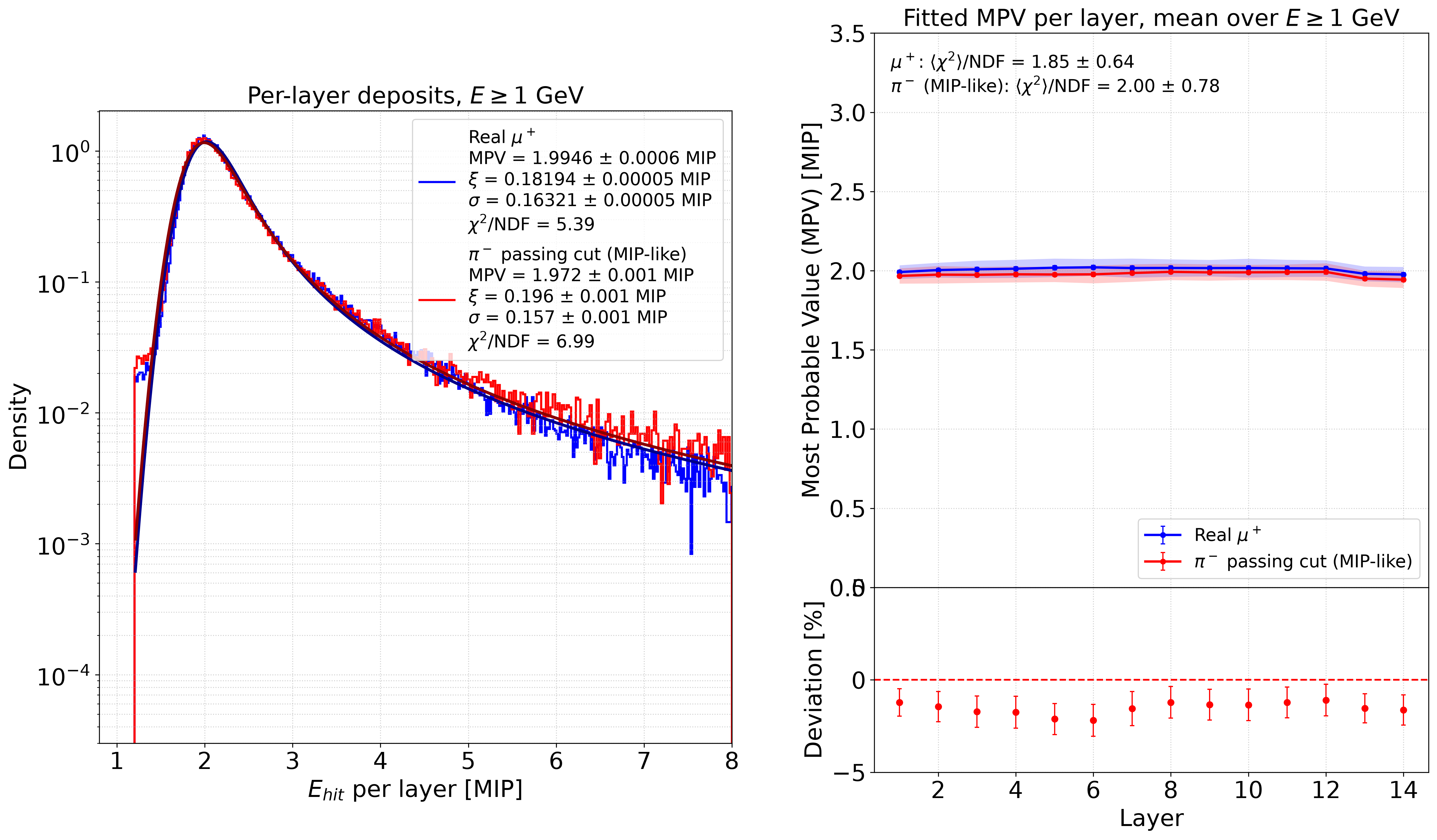}
    \caption{Left: Landau-Gaussian fits to the energy distribution for all bars with energies greater than $\qty{1}{\giga \electronvolt}$. Solid and dashed lines indicate simulation and fits, respectively. Right: Per-layer MPV values of fits with particle energies.  Shaded areas indicate the standard deviation of the MPV across particle energies from \qtyrange{1}{10}{\giga \electronvolt}. The bottom subplot shows the ratio between the $\mu^{+}$ and $\pi^{-}$ MPV, indicated by the red dots. Blue and red lines and shaded regions in both plots represent the $\mu^{+}$ and $\pi^{-}$ respectively. Fit values and the mean and standard deviation of the reduced-$\chi^{2}$ values are shown in the legend.}
    \label{fig:fig_landau_mpv_mu+_vs_pi-}
\end{figure}

The total distribution of energies and the fitted MPV per layer for both classes are shown in Fig.~\ref{fig:fig_landau_mpv_mu+_vs_pi-}. Example fits per particle energy are shown in Appendix Fig.~\ref{fig:fig_landau_fit}. 
The results show that $\mu^{+}$ and MIP-like $\pi^{-}$ events follow a Landau-Gaussian distribution with good agreement, with reduced $\chi^{2}$ values of 5.39 and 6.99 in both cases. When fitted layer by layer, the agreement improves to reduced $\chi^{2}$ values of \num{1.85(0.64)} and \num{2.0(0.78)}, respectively. The energy deposited per layer in both cases is nearly a constant value of \qty{2.0}{\mathrm{MIP}}, corresponding to the sum of two SiPMs independently calibrated to \qty{1}{\mathrm{MIP}} for each bar. The MPV values differ by only \qty{1.53(0.26)}{\percent} over the layers. This result can be explained by the marginally greater MPV of $\mu^{+}$ than $\pi^{-}$ owing to its smaller mass. This is illustrated in Appendix Fig.~\ref{fig:fig_bethe_bloch_beta2_region}, which shows the differential energy loss per layer for $\mu^{+}$ and $\pi^{-}$, for which the loss is greater for $\mu^{+}$ for energies above around \qty{0.5}{\giga \electronvolt}. It is noted that this MPV value is averaged across several energies on which there is a small expected dependence, which is not shown. 

The events failing the cut were expected to be showering $\pi^{-}$ events. By fitting the expected longitudinal hadronic shower shape event by event, the shower start position, $z_{\mathrm{ss}}$, can be extracted and studied. The model for the shower shape is described as a mixture of a uniform distribution to describe the MIP track prior to the shower, and a Gamma distribution for the showering particle, which is often used to parameterize the longitudinal development of particle showers \cite{bock_parametrization_1981, 
longo_monte_1975,
chadeeva_parametrisation_2015, grindhammer_parameterized_2000, pinto_shower_2022} 

\begin{equation}
    \diff{E}{z} = E_{\mathrm{MIP}} \cdot  \operatorname{Uniform}(z;\, 0, z_{\mathrm{ss}})
    + E_{\mathrm{shower}} \cdot \operatorname{Gamma}(z;\, \alpha_{z}, \beta_{z}, \Delta_{z} = z_{\mathrm{ss}}),
    \label{eq:profile_model}
\end{equation}

\begin{equation}
    \operatorname{Gamma}(z;\, \alpha_{z}, \beta_{z}, \Delta_{z}) =
    \frac{\beta_{z}^{\alpha_{z}}}{\Gamma(\alpha_{z})} \,
    (z - \Delta_{z})^{\alpha_{z} - 1} \, e^{-\beta_{z} (z - \Delta_{z})} \,
    \Theta(z - \Delta_{z}),
    \label{eq:gamma_pdf}
\end{equation}

\begin{equation}
    \operatorname{Uniform}(z;\, a, b) = \frac{1}{b - a} \, \Theta(z - a) \, \Theta(b - z).
    \label{eq:uniform_pdf}
\end{equation}

where $\beta$ is a shower rate parameter, $\beta = (\alpha-1)/T_{z}$, $T_{z}$ is the shower's maximum position, $\alpha_{z}$ is a shape parameter that characterizes the shower's symmetry, and $\Delta_{z}$ is an offset. $E_{\mathrm{MIP}}$ and $E_{\mathrm{shower}}$, are the energies deposited by the ionizing track and shower, respectively, with total energy $E_{\mathrm{sum}} = E_{\mathrm{MIP}} + E_{\mathrm{shower}}$. Two physical bounds are imposed: the MIP plateau
height is restricted to $0.3 \leq E_{\mathrm{MIP}} / z_{\mathrm{ss}} \leq
8$ MIP per layer, which constrains the total energy of the MIP track prior to the shower to Landau-distributed energies. The shower rate parameter is the reciprocal of the scale in $z$, and is therefore constrained such that the shower scale is physical, $\beta_z \geq \frac{1}{3}$ layer$^{-1}$. In addition, the shower start is required to precede the shower maximum, $z_\mathrm{ss} < T_{z}$. The distributions were fitted using the \texttt{MIGRAD} algorithm of \texttt{iminuit} with an extended binned log-likelihood, under the simplifying assumption that the bin fluctuations are Poisson-distributed. Weighting methods such as the Bohm-Zech transformation, which uses an effective Poisson count $n_{\mathrm{eff}} = (\Sigma E_{i})^2 / \Sigma E_{i}^2$, were considered. However, in low-occupancy bins where only a couple of high-energy hits dominate, $n_{\mathrm{eff}}$ becomes small, and the fit becomes unstable. Low-energy bins must nonetheless be included due to the limited number of bins available. However, the unweighted Poisson assumption is incorrect and underestimates the uncertainty in each bin, resulting in cases where reduced-$\chi^{2}$ can be less than 1. Therefore, to interpret the results, the Wasserstein-1 ($W_{1}$) statistic is used to assess the fit between the empirical cumulative distribution function (ECDF) of the data and the binned CDF. The Wasserstein-1 statistic, or 'Earth-Mover's Distance', is used to describe the fit quality. The statistic indicates the average 'distance' required to transport the probability mass from one distribution to the other. It is defined as the integral over the absolute distance in probability between two CDFs:

\begin{equation}
    W_{1} = \int^{\infty}_{-\infty} \left| F(x) - G(x) \right| 
\end{equation}

where $F$ and $G$ are two CDFs. 

\begin{figure}
    \centering
    \includegraphics[width=\linewidth]{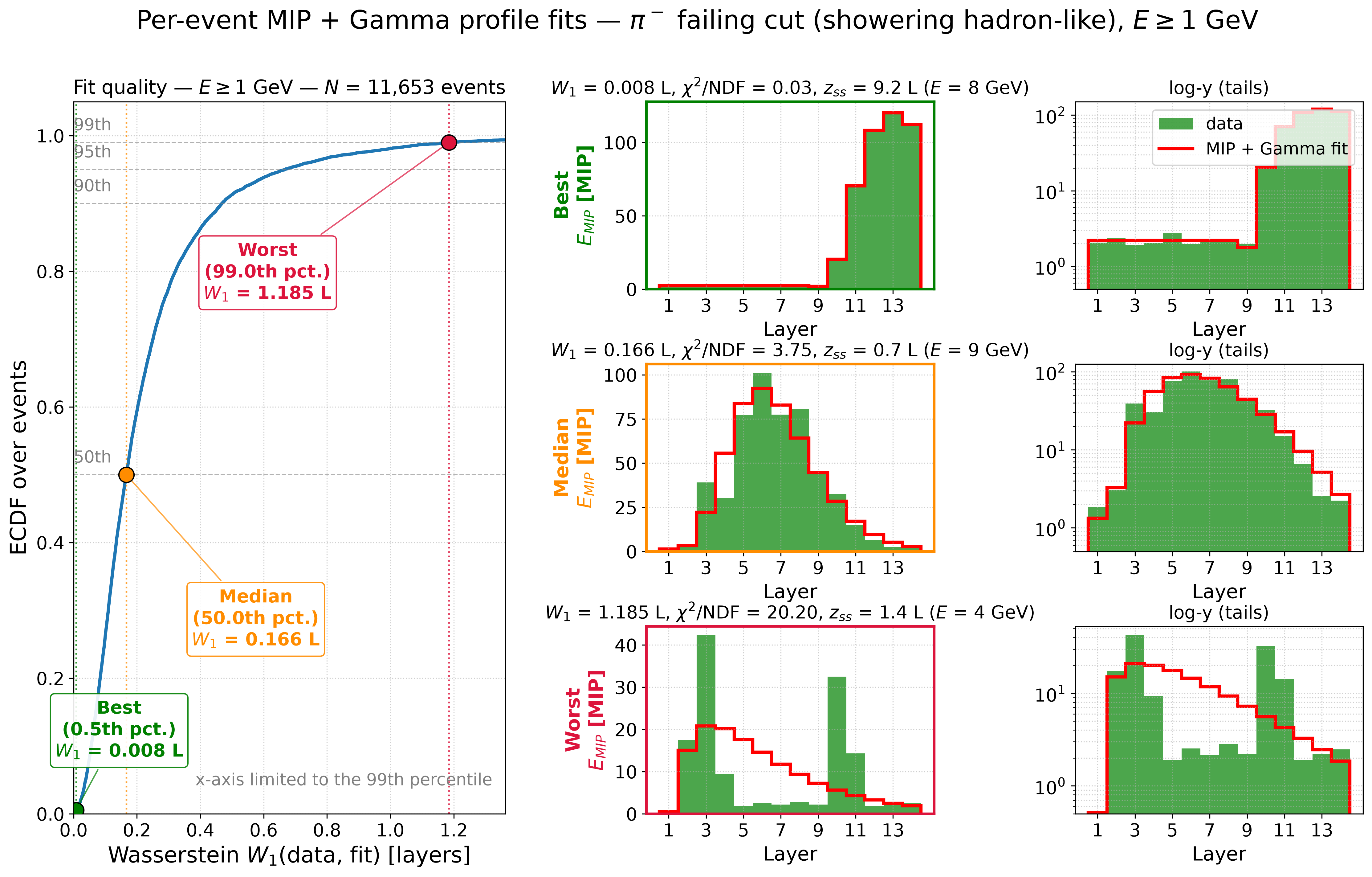}
    \caption{Examples of the best, median and worst fits for $\pi^{-}$ events failing the cut with energy greater than \qty{1}{\giga \electronvolt}. The left subfigure shows the empirical cumulative distribution function (ECDF) vs. the Wasserstein-1 values, $W_1$, and the percentiles shown on the right in grey text. The green, blue, and red text indicate the best, typical (median), and worst case for the fits up to the \qty{99}{\percent} percentile. The green energy-weighted histograms are shown on the right, with each row corresponding to the colored axis label. The $y$-axis of the histograms shows the energy deposited per bar index or layer. The red line indicates the fitted profile for that event.}
    \label{fig:fig_shower_start_example_fits_mu+_vs_pi-}
\end{figure}

The results are shown in Fig.~\ref{fig:fig_shower_start_example_fits_mu+_vs_pi-}. These show that the events failing the cut are generally well described as showering particles with ionizing tracks, corresponding to hadron showers. It is noted that the worst-fitting example shown includes multiple subshowers, which could be fitted with additional Gamma terms. 

The pion interaction length, $\lambda_{\pi}$, was then fitted using the $z_{\mathrm{ss}}$ value under the assumption of an exponential distribution.  The truncated exponential distribution was fitted using Eq.~\ref{eq:truncated_exp}:

\begin{equation}
    f(z_{\mathrm{ss}}; \lambda_{\pi}) =
    \frac{e^{-\frac{(z_{\mathrm{ss}} - z_{\min})}{\lambda_{\pi}}}}
         {\lambda_{\pi} \left( 1 - e^{-\frac{(z_{\max} - z_{\min})}{\lambda_{\pi}}} \right)},
    \qquad z_{\min} \le z_{\mathrm{ss}} \le z_{\max}
    \label{eq:truncated_exp}
\end{equation}

where $z_{\min}=\qty{2}{\mathrm{layer}}$ and $z_{\max}=\qty{10}{\mathrm{layer}}$. 

The reason for the truncation and restricted fit range was that a skewed subset of pathological fits was observed outside this range. Failed fits were observed in the range $z_{\mathrm{ss}}<2$, where the MIP and shower components become nearly degenerate. In these cases, the Gamma shower profile absorbed the MIP track. Additionally, cases with multiple subshowers were also observed, as shown in Fig.~\ref{fig:fig_shower_start_example_fits_mu+_vs_pi-}. In the range $z_{\mathrm{ss}}>10$, only three layers remain, which were observed to be insufficient to fit the showering component.

\begin{figure}
    \centering
    \includegraphics[width=0.5\linewidth]{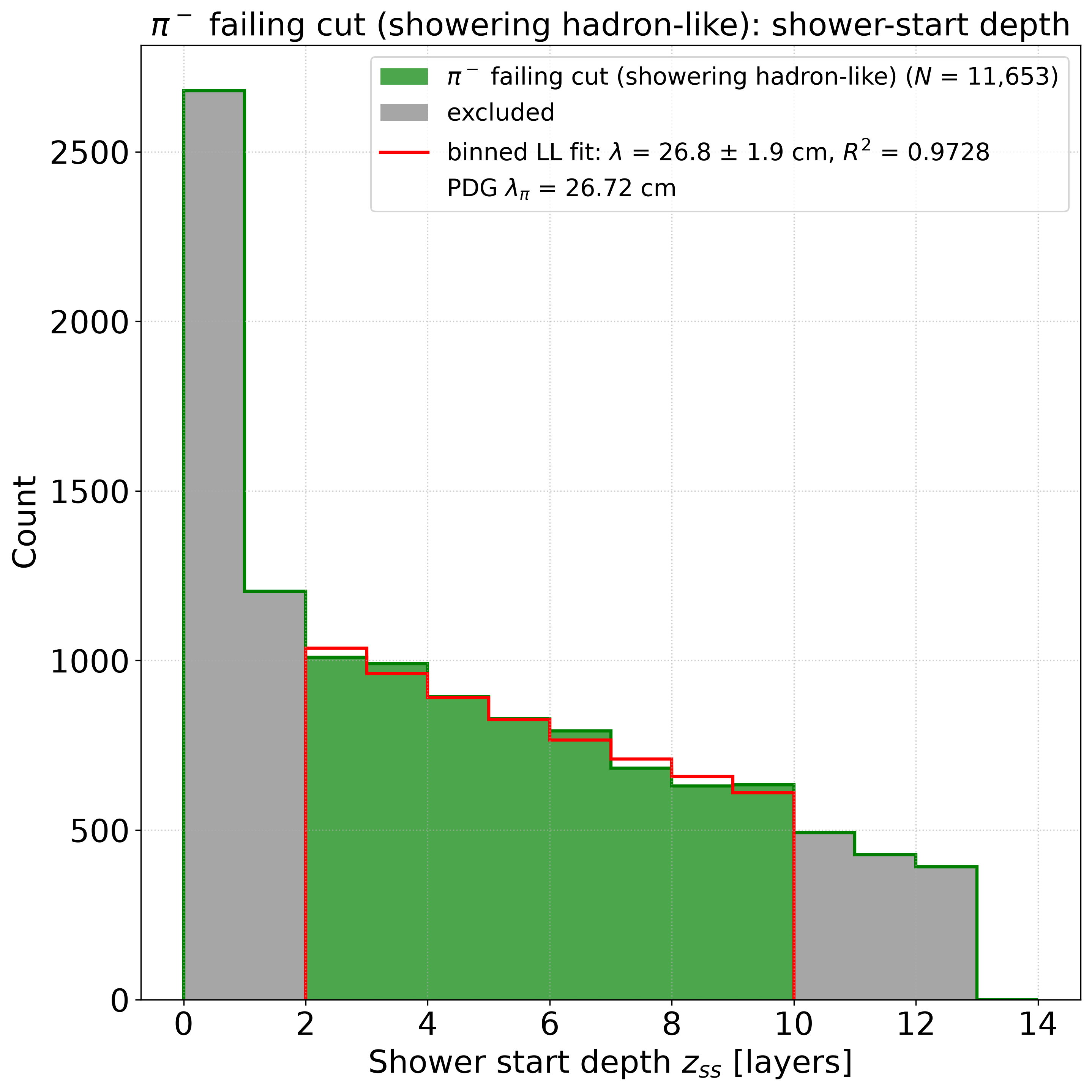}
    \caption{Fit of the pion interaction length from the fitted shower start positions of $\pi^{-}$ events that fail the cut. The green and grey shaded areas represent the fitted and excluded regions, respectively. The red line indicates the truncated exponential fit. Fit values and reduced $\chi^{2}$ are shown in the legend.}
    \label{fig:fig_shower_start_mu+_vs_pi-}
\end{figure}

The results are shown in Fig.~\ref{fig:fig_shower_start_mu+_vs_pi-}. The fitted pion interaction length was found to be \qty{26.8(1.9)}{\centi \meter}. This matches the value quoted by the PDG of \qty{26.72}{\centi \meter} from Refs.~\cite{pdg_review_2026, pdg_review_2026_url}. It may therefore be argued that the showering particles are well described according to expectations of showering $\pi^{-}$ events.

SHapley Additive exPlanations (SHAP) \cite{lundberg_local_2020} is a model-agnostic technique for interpreting machine learning models, rooted in game-theoretic Shapley values \cite{shapley_notes_1951}. This method quantifies the significance of each input variable to the model's decision-making process by measuring the average marginal effect of including that variable across all subsets of the other features. This can be achieved in practice by replacing the variable with random values from a background distribution, which approximates 'removing' that feature from the model and is similar in spirit to the 'leave-one-out' method of jackknife re-sampling. and studying the deviation of the output of the model compared to its average output. 

The graph neural network classifier output, given the $\mu^{+}$ prototype, can be studied using the \texttt{Captum} library with the \texttt{ShapleyValueSampling} method \cite{kokhlikyan_captum_2020, noauthor_meta-pytorchcaptum_2026}. The result is a 'SHAP value': a positive value promotes the selection of that class, and a negative value discourages it. This can be used to study how the model makes a decision based on the input. 

\begin{figure}
    \centering
    \includegraphics[width=0.7\linewidth]{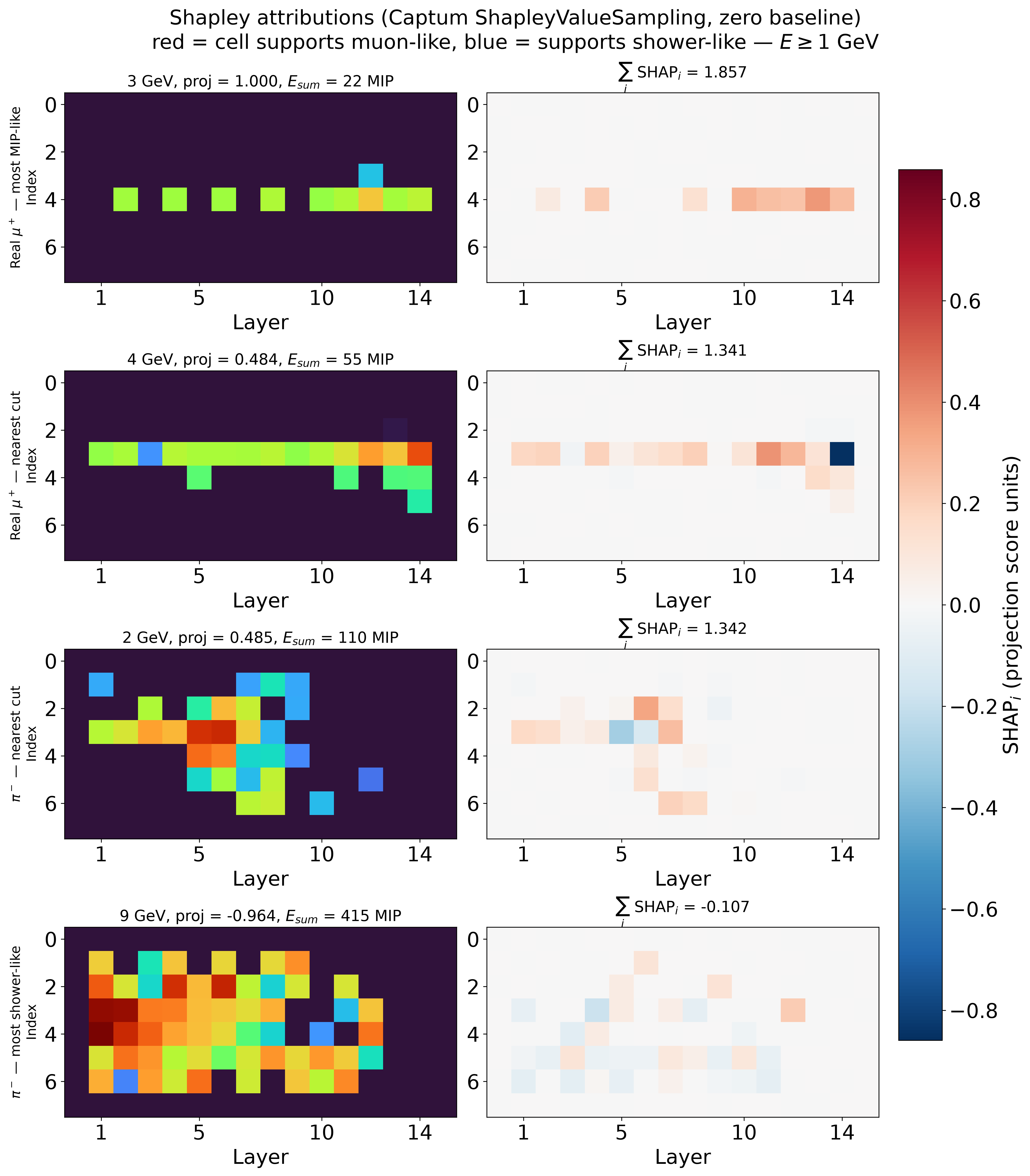}
    \caption{Example two-dimensional event displays of $\mu^{+}$ and $\pi^{-}$ with corresponding SHAP values. The top/second row ($\mu^{+}$) and the third/bottom row ($\pi^{-}$) show the example events at the extremes of the cut (top and bottom row) and the closest examples to the cut boundary (second and third row). The left column shows the event displays, with blue to red color indicating low to high energy deposited in each bar on a log scale.  The projection value on the prototype and the energy sum are shown in the title. The right column shows the corresponding SHAP values for each active bar, with blue to red color indicating negative to positive promotion of the $\mu^{+}$ class by that bar. In both columns, the $x$ and $y$ correspond to the lateral index and longitudinal layer, respectively.  
    }
    \label{fig:fig_event_displays_shap_mup_vs_pim}
\end{figure}

Event displays and their corresponding SHAP values per bar are shown in Fig.~\ref{fig:fig_event_displays_shap_mup_vs_pim}. It can be observed that negative inference bars typically have higher energies than those bars with positive inference, without an observably strong correlation to the spatial position. It was predicted that the optimized cut worked primarily to classify using the bar-by-bar energy distribution rather than its distribution in space. 

To assess this hypothesis, the joint distribution of the average SHAP value per bar was plotted against the expectation value of the log-likelihood of the energy coming from the Landau-Gaussian distribution fitted in Fig.~\ref{fig:fig_landau_mpv_mu+_vs_pi-}, with fitted parameters for the real $\mu^{+}$ data, $\mathbb{E}(\ln{P(E_{\mathrm{bar}}} | \operatorname{LandauGaus}(x; \mu, \sigma, \text{MPV}, \xi) ))$. The SHAP value and   mean log probability are calculated for each event. 

The mean SHAP value is a  measure of the total vote for the MIP-like class given every bar in the event. If the measured probability can act as a sufficiently accurate classifier that performs similarly to the unsupervised cut, it may be inferred that the neural network has learned this property as the main discriminating feature between the two classes. Therefore, a 1D linear discriminant analysis (i.e a single threshold value) was fitted to maximally separate the classes using the expected log-likelihood statistic. The AUC of the ROC curve obtained from this classifier indicates its similarity to the cut that was optimized by the Siamese neural network.

\begin{figure}
    \centering
    \includegraphics[width=0.6\linewidth]{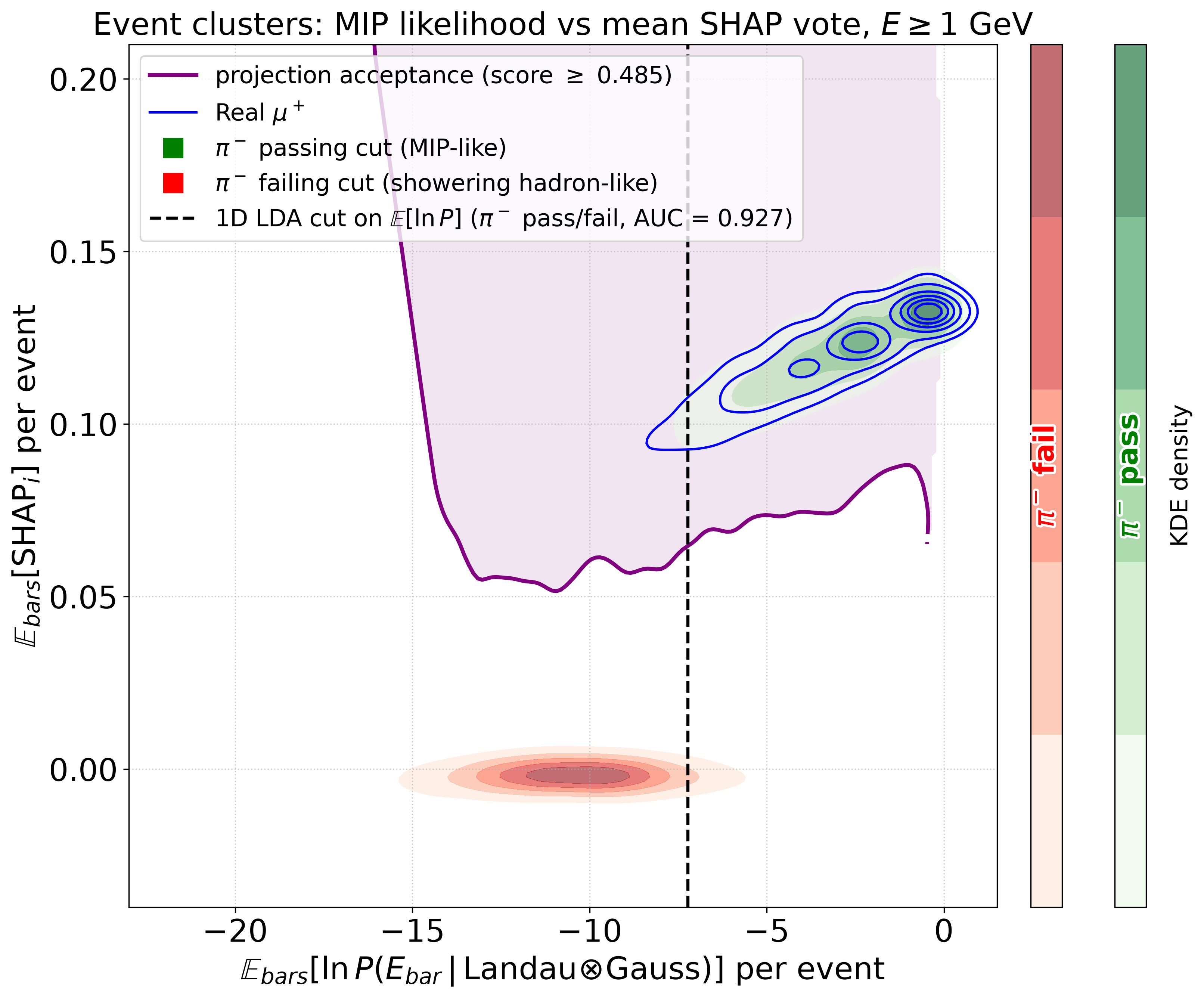}
    \caption{Joint distribution of the average SHAP value of a bar per event and the average log-likelihood of the bar energy coming from a Landau-Gaussian distribution. The green and red kernel density estimates (KDE) show the smoothed probability distributions of the $\pi^{-}$ events passing and failing the cuts, respectively. The blue line contours indicate the distribution of the $\mu^{+}$ events passing the cut. The dashed line indicates the 1D linear discriminator between the two classes of $\pi^{-}$ events. The purple region indicates the decision acceptance boundary for MIP-like events.}
    \label{fig:fig_event_clusters_mu+_vs_pi-.png}
\end{figure}

The results are shown in Fig.~\ref{fig:fig_event_clusters_mu+_vs_pi-.png}. The probability density clusters of events passing and failing the cut are clearly separated in both the expectation value of the SHAP value and the expectation value of the log-likelihood of the energy deposited given the fitted Landau-Gaussian distribution. The linear discriminant classifier using the expected value of the log-likelihood per event achieves an AUC between MIP-like and showering hadron-like, as defined by the unsupervised classifier, of 0.927. This study illustrates that the tag can generally be understood from physical expectations that on average, the

\begin{figure}
    \centering
    \includegraphics[width=0.6\linewidth]{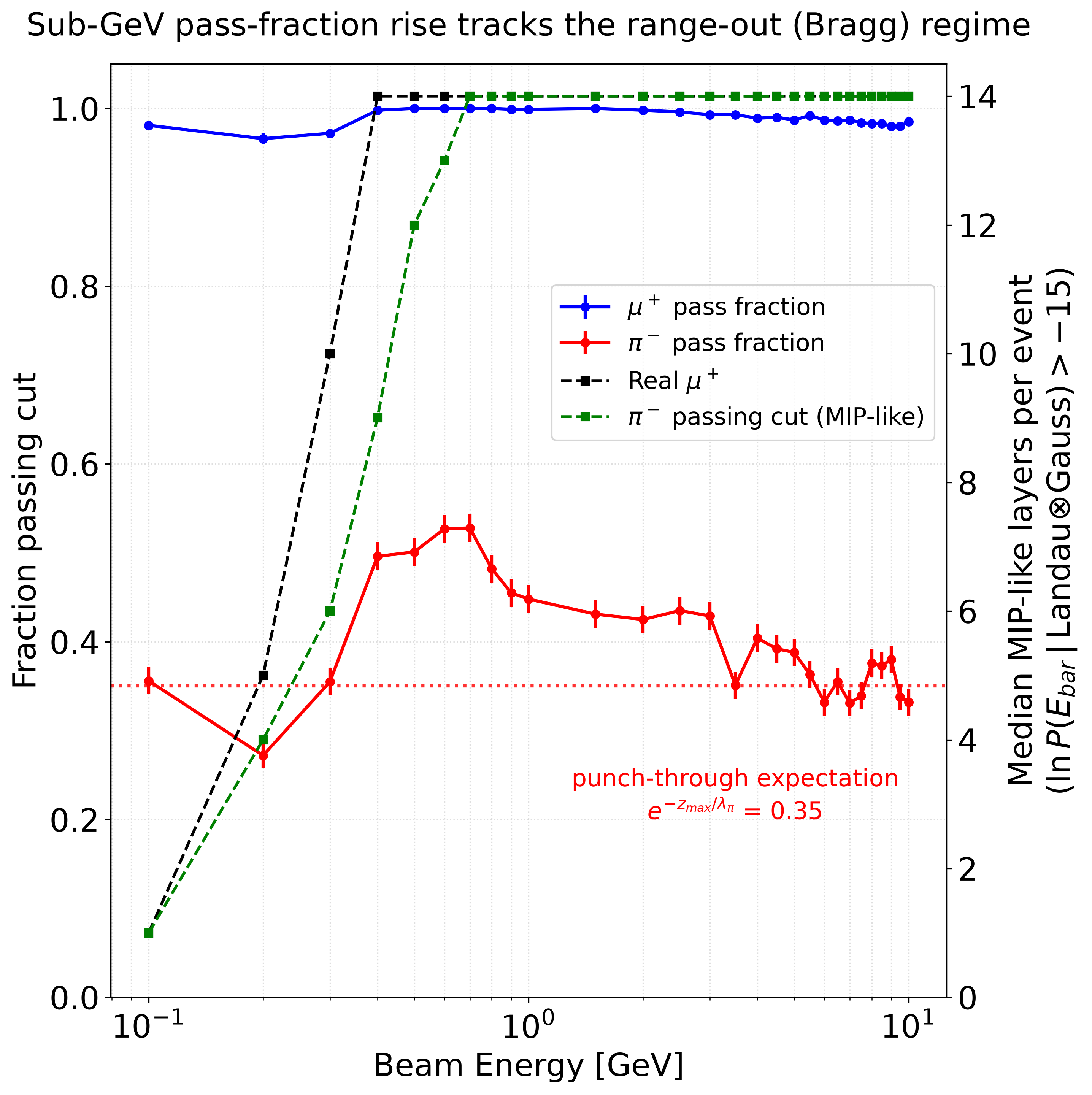}
    \caption {Punch-through fraction and median MIP-like layers per event with the log-likelihood of the energy being greater than -15. The blue and red lines show the fractions of the $\mu^{+}$ and $\pi^{-}$ data passing the cut. The dashed red line indicates the expected average punch-through pion fraction. The dashed black and green lines indicate the number of MIP-like layers for $\mu^{+}$ and $\pi^{-}$, respectively.}
    \label{fig:fig_event_clusters_mu+_vs_pi-}
\end{figure}

Finally, the fraction of events accepted in the simulation by the cut is shown in Fig.~\ref{fig:fig_event_clusters_mu+_vs_pi-}. For the $\pi^{-}$ sample, the fraction is compared to the punch-through fraction, defined as the median number of layers per event for which the median log-likelihood of the energy deposited per layer being drawn from the Landau distribution is greater than -15. The plot shows that the fraction passing the cut is close to 1.0 for $\mu^{+}$ and at energies above around \qty{2}{\giga \electronvolt}, close to the expected punch-through probability of 0.35. Below \qty{1}{\giga \electronvolt}, the fraction starts to increase. The increase is observed most strongly where $\pi^{-}$ are observed to begin to range out in the calorimeter. Appendix Fig.~\ref{fig:fig_bethe_bloch_beta2_region} compares the Bethe-Bloch energy loss for $\pi^{-}$ and $\mu^{+}$ for different initial particle energies. It shows that for $\pi^{-}$, the Bragg region occurs at greater incident momenta than for $\mu^{+}$ due to the latter's higher mass.  

In summary, the unsupervised method is demonstrated to be able to perform successfully as a classifier for $\mu^{+}$ and $e^{-}$ events without labels, and similarly distinguish punch-through from showering $\pi^{-}$ by comparison with $\mu^{+}$. The method allows the re-derivation of an approximately equivalent cut using the expectation values of the log-likelihood of the Landau-Gauss distribution over all hits. This cut was identified through SHAP analysis. The method assumed no prior knowledge of the events before classification.

\subsection{Supervised PID}\label{sec:SupervisedPIDResults}

\begin{figure}
\centering

\begin{subfigure}{0.65\linewidth}
\centering
\includegraphics[width=\linewidth]{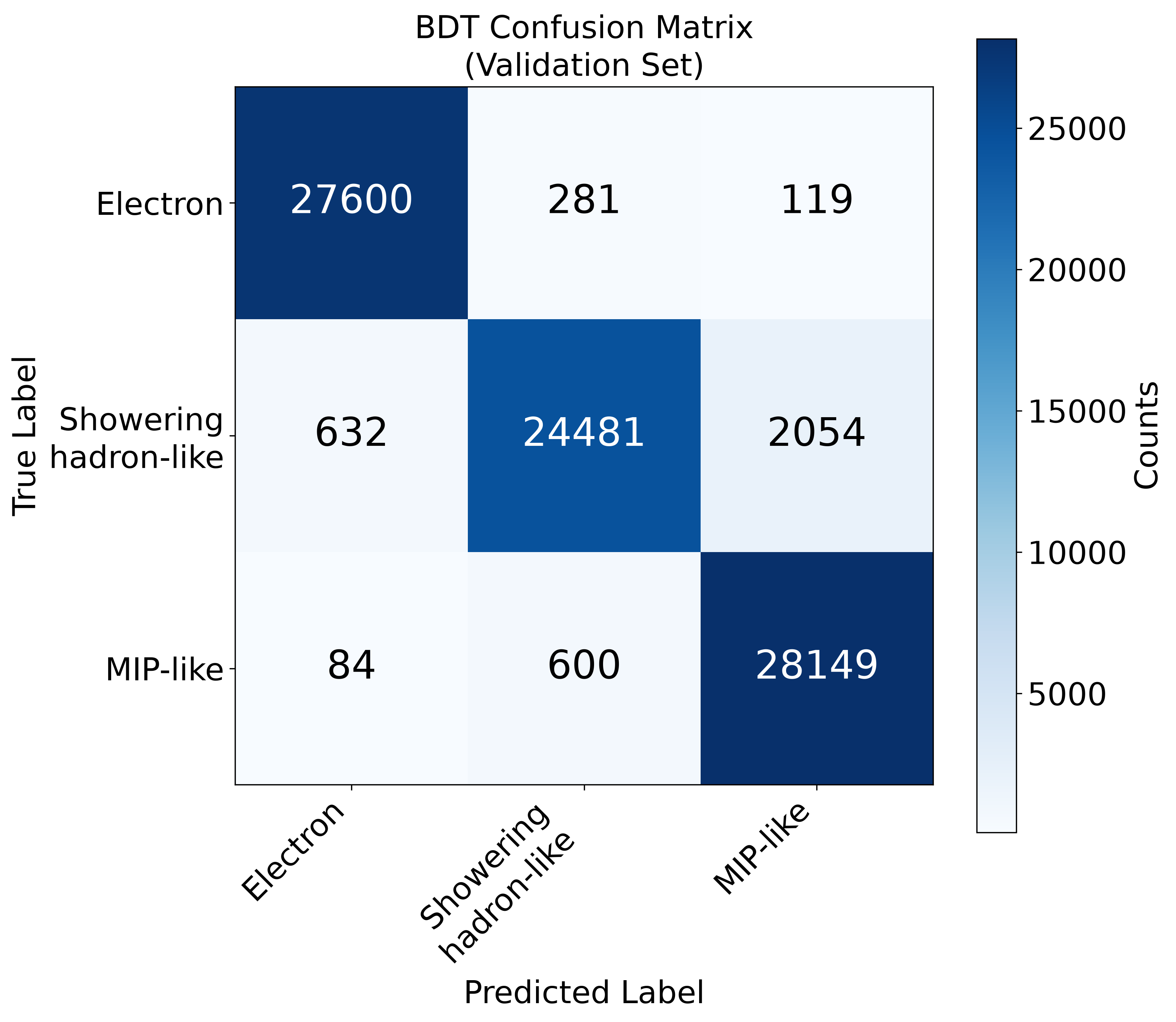}
\caption{}
\label{fig:fig_BDT_confusion_matrix}
\end{subfigure}

\vfill

\begin{subfigure}{0.49\linewidth}
\centering
\includegraphics[width=\linewidth]{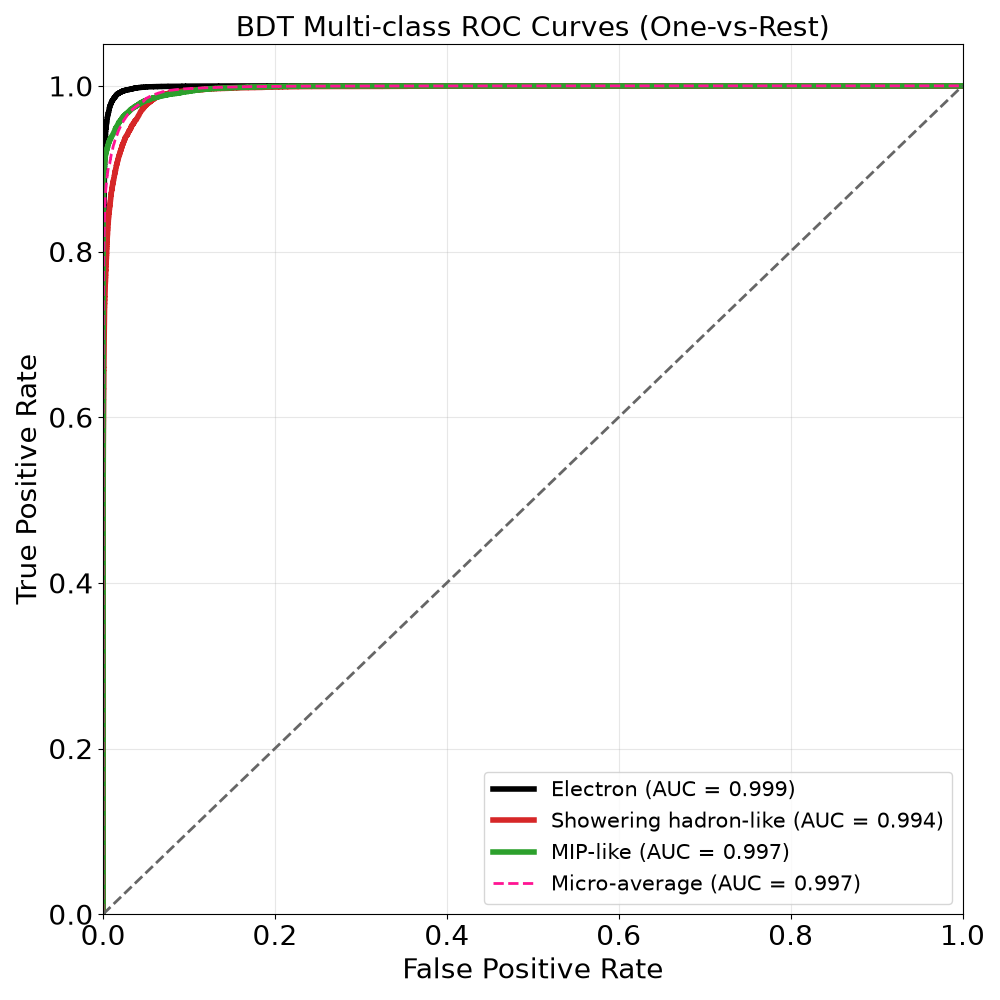}
\caption{}
\label{fig:fig_BDT_roc}
\end{subfigure}
\hfill
\begin{subfigure}{0.49\linewidth}
\centering
\includegraphics[width=\linewidth]{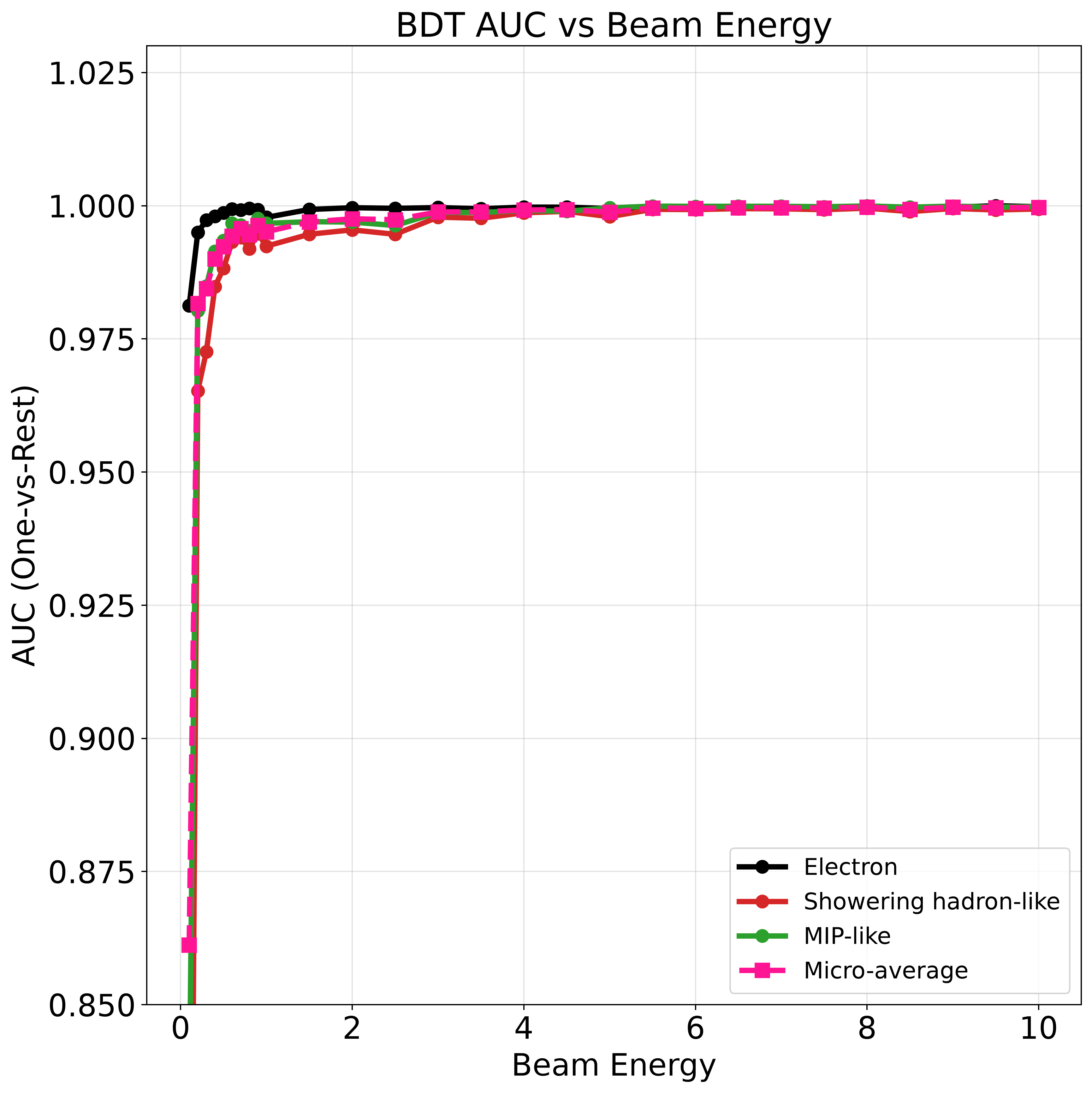}
\caption{}
\label{fig:fig_BDT_auc_vs_beamEnergy}
\end{subfigure}

\caption{Performance metrics of the classification of electron, showering hadron and MIP-like events for the BDT. Fig.~\ref{fig:fig_BDT_confusion_matrix} shows the confusion matrix for the BDT classifier. Fig.~\ref{fig:fig_BDT_roc} shows the ROC curve, where the black, green and red lines correspond to the ROC curve for the electron, showering hadron and MIP-like classification vs. all other classes. The red dashed line indicates the micro-average (overall ROC curve for all classes against all other classes). The legend shows the AUC for each case. Fig.~\ref{fig:fig_BDT_auc_vs_beamEnergy} shows the performance of the classifier in terms of the AUC of one class vs all others (one vs. rest) and micro-average as a function of the particle energy, with the same colour scheme as Fig.~\ref{fig:fig_BDT_auc_vs_beamEnergy}. }
\label{fig:fig_BDT_results}
\end{figure}

\begin{figure}
\centering

\begin{subfigure}{0.65\linewidth}
\centering
\includegraphics[width=\linewidth]{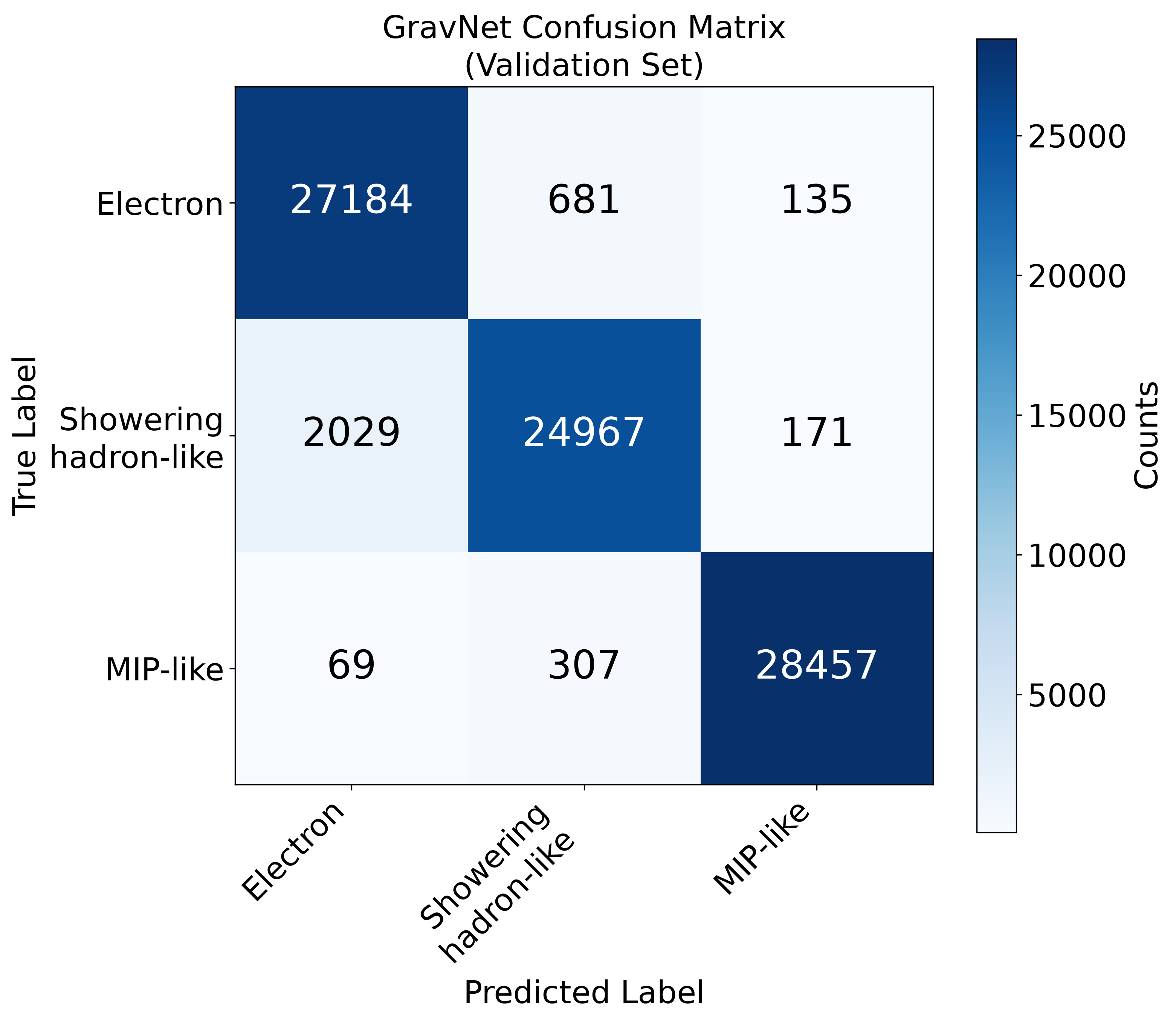}
\caption{}
\label{fig:fig_GravNet_confusion_matrix}
\end{subfigure}

\vfill

\begin{subfigure}{0.49\linewidth}
\centering
\includegraphics[width=\linewidth]{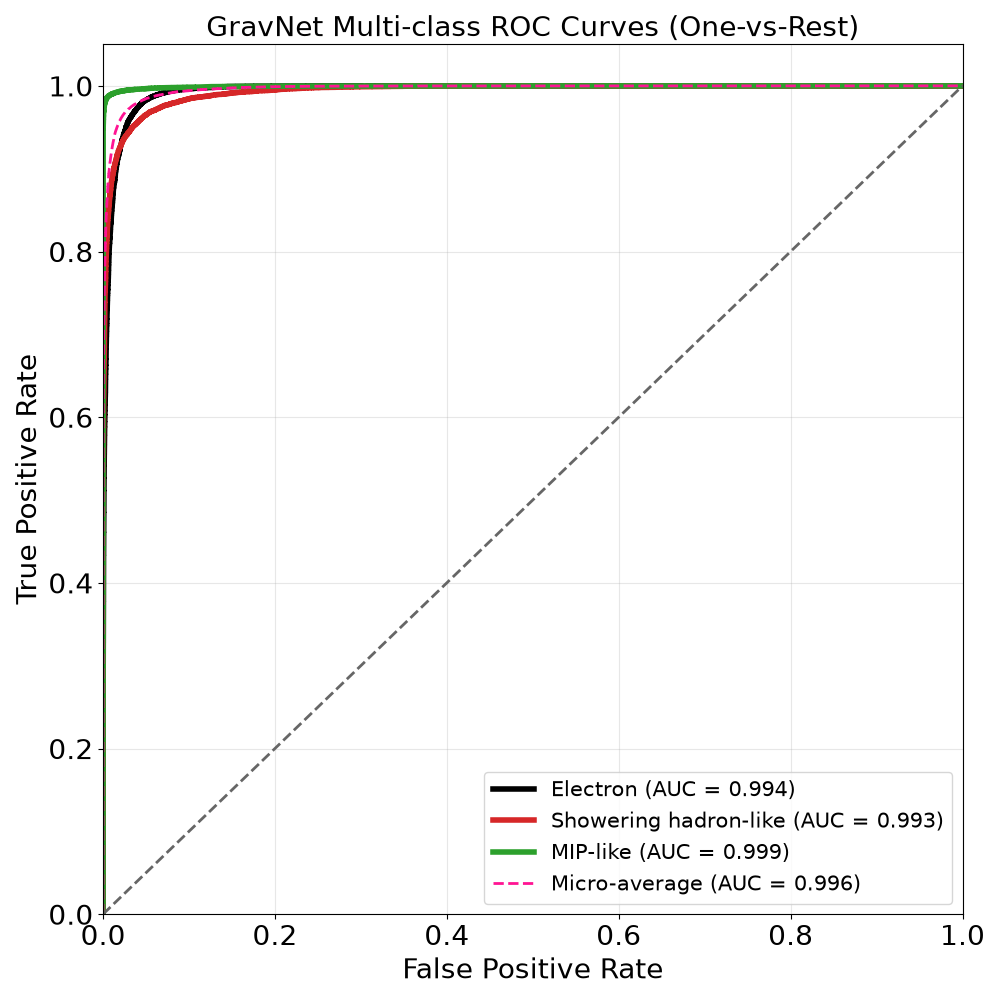}
\caption{}
\label{fig:fig_GravNet_roc}
\end{subfigure}
\hfill
\begin{subfigure}{0.49\linewidth}
\centering
\includegraphics[width=\linewidth]{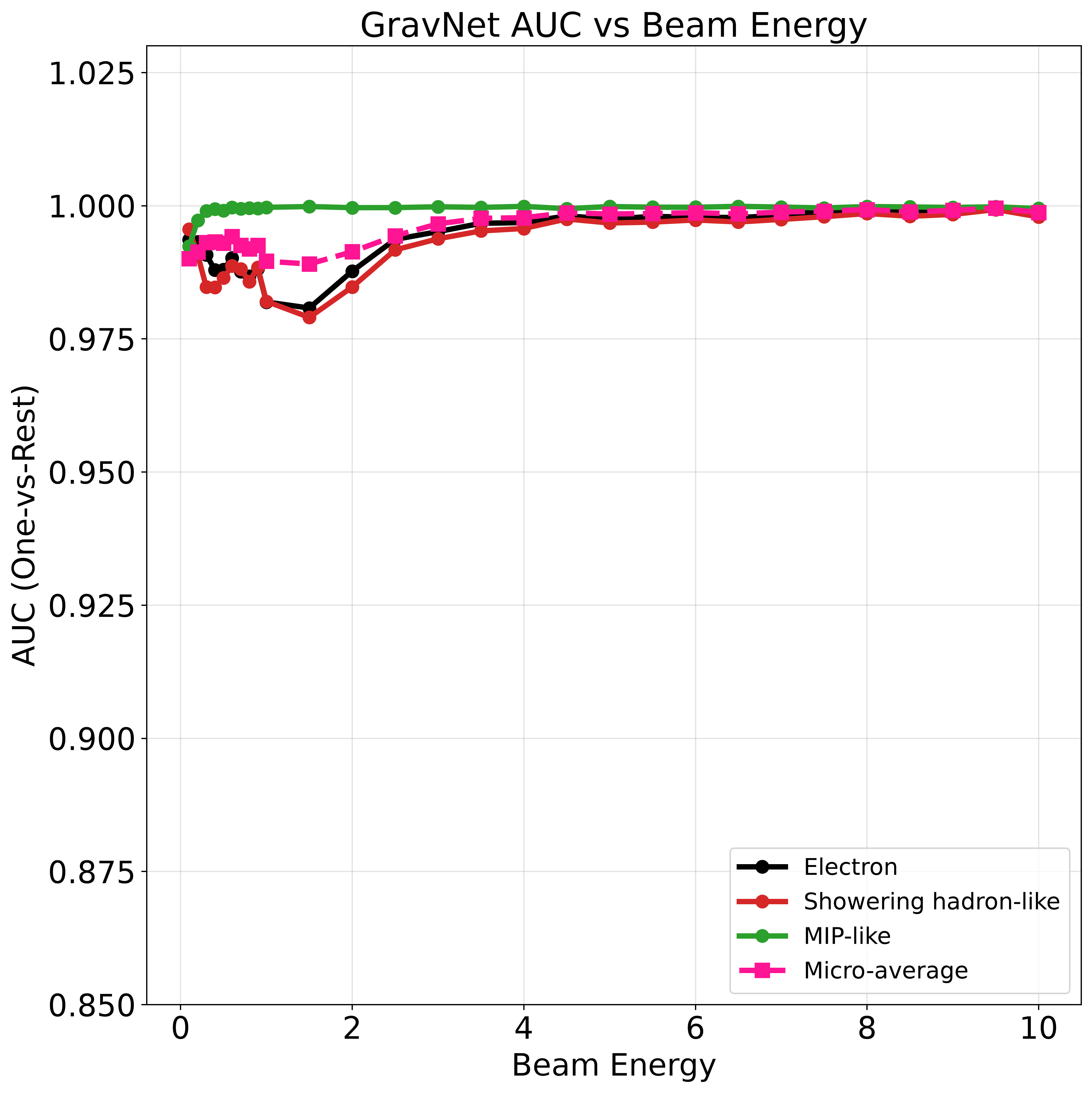}
\caption{}
\label{fig:fig_GravNet_auc_vs_beamEnergy}
\end{subfigure}

\caption{Performance metrics of the classification of electron, showering hadron and MIP-like events for the \texttt{GravNet} model. Otherwise, the figures correspond to the descriptions in Fig.~\ref{fig:fig_BDT_results}. }
\label{fig:fig_GravNet_results}
\end{figure}

The confusion matrices, ROC curves, and AUCs as a function of particle energy for the supervised models to separate 'electron', 'showering hadron ', and 'MIP-like' event classes as defined earlier are presented for \texttt{GravNet} and BDT models in Figs. \ref{fig:fig_BDT_results} and \ref{fig:fig_GravNet_results}, respectively. 

In both cases, a total micro-average AUC, defined as the AUC comparing the performance across all classes, was found to be 0.996. This indicates highly effective clustering for both models. The fact that both models achieved this result independently suggests that the calorimeter captures sufficient energy density information that the event classes can be trivially identified with high accuracy using the crystal ECAL prototype.

The comparison of Fig.~ \ref{fig:fig_BDT_confusion_matrix} and Fig.~ \ref{fig:fig_GravNet_confusion_matrix} reveals that the classification performance varies between model types. The BDT model is more accurate at determining electrons but is less accurate at distinguishing MIP-like particles from showering hadrons. The \texttt{GravNet} model is more accurate at identifying MIP-like particles but less accurate at distinguishing showering hadrons from electrons. In summary, distinguishing showering hadrons is the most challenging task for either model. This is also shown in Fig.~\ref{fig:fig_BDT_confusion_matrix}, where the one vs. all AUC scores for electrons, showering hadrons, and MIP-like particles are $\left[0.999, 0.994, 0.997\right]$ for the BDT model and $\left[0.994, 0.993, 0.999\right]$ for the \texttt{GravNet} model, respectively. 
This result can be understood as a consequence of the model design. Graph neural networks are effective at describing  sparsity between nodes, aiding identification of the regular structure of MIP-like events. By contrast, a BDT uses aggregated event information, which is more informative for events where there are many individual energy deposits such as electrons. It is noted that an ensemble using both methods would likely improve the performance overall, though this is left to a future study.

The AUC of the classifier as a function of particle energy is shown in Fig.~\ref{fig:fig_BDT_auc_vs_beamEnergy} and Fig.~\ref{fig:fig_GravNet_auc_vs_beamEnergy}. For particle energies below \qty{2}{\giga \electronvolt}, both models degrade in performance, and above \qty{2}{\giga \electronvolt} they saturate at close to 1.0. For the BDT model, performance sharply degrades as the model approaches \qty{0.1}{\giga \electronvolt}, whereas for the \texttt{GravNet} model, the performance in this region is worse overall but more consistent in the very low energy regime.

The boosted decision tree enables a detailed exploration of how the classifier performs and indicates what information is useful for the model in a particular context. The model's performance can be related back to the physics of particle showers.

\begin{figure}
    \centering
    \includegraphics[width=\linewidth]{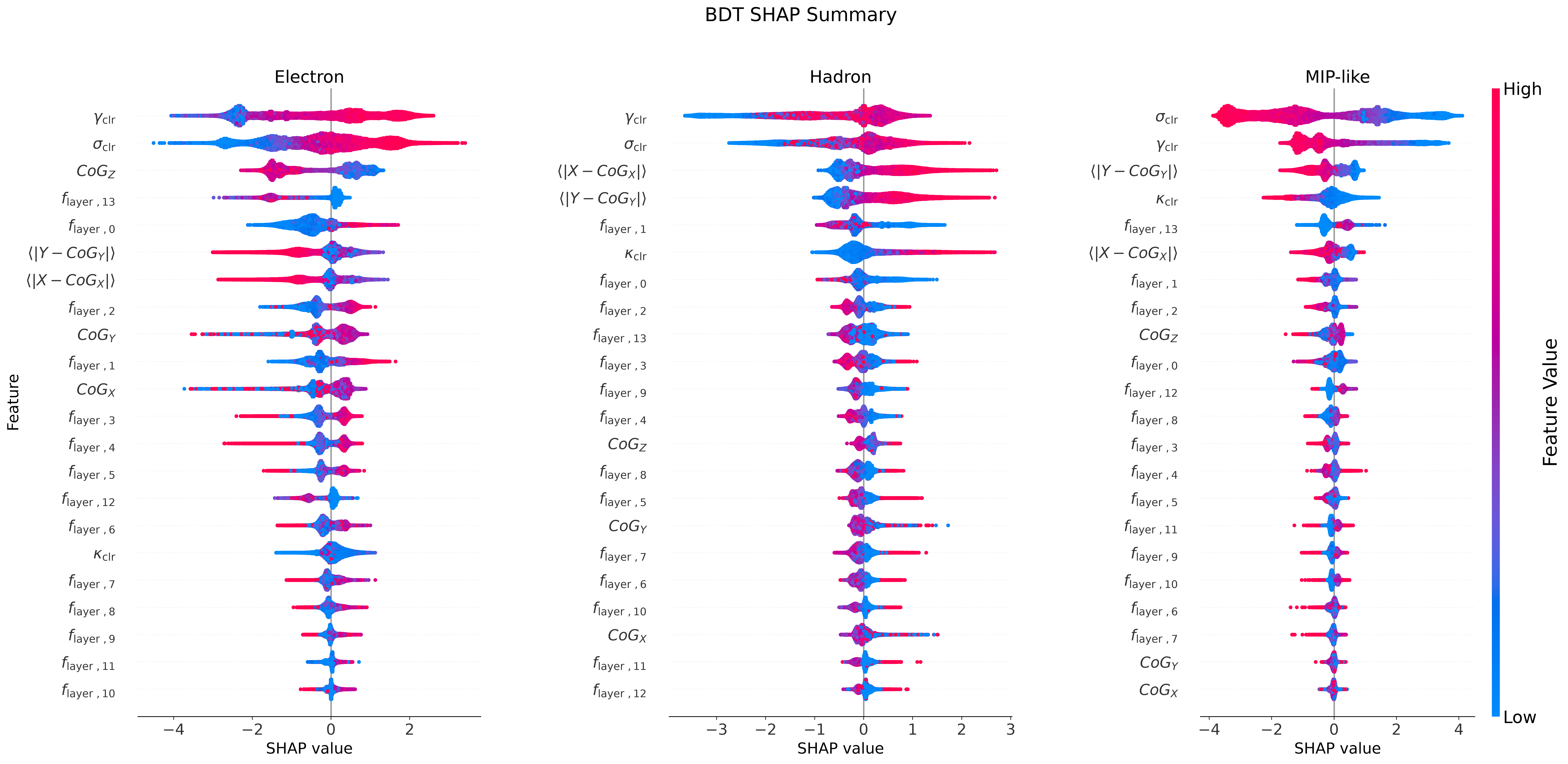}
    \caption{SHAP 'beeswarm' plot for the BDT model, showing the relationship between the input variable value and the influence of the variable on the model prediction. Each plot, left to right, shows the electron, showering hadron, and MIP-like event classes. In each case, the variables are ranked from top to bottom according to their influence on the model's predictions. A negative to positive SHAP value indicates negative to positive influence of the variable on the model's final decision. The blue-to-pink color scheme indicates the scale of the parameter, from the 5\textsuperscript{th} to the 95\textsuperscript{th} percentile (small-to-large).} 
    \label{fig:fig_BDT_shap}
\end{figure}

A SHAP analysis was performed using the \texttt{TreeExplainer} in the eponymous \texttt{SHAP} software library \cite{noauthor_shapshap_2026}, which can compute exact SHAP values efficiently for tree-based models.
A 'beeswarm' plot of the BDT SHAP-values with ranked variables produced from the \texttt{TreeExplainer} is shown for each class in Fig.~\ref{fig:fig_BDT_shap}. The plot can be interpreted as follows:
\begin{itemize}
    \item Electrons are selected using a small longitudinal center of gravity and average distance from the lateral center of gravity, little energy deposited in the final layer of the calorimeter, and a broad skewness and spread of the central log ratio of the bar energies within the particle shower. The model therefore selects energy-compact, shallow particle showers that shower immediately to be electrons; 

    \item  Showering hadron-like events are selected using a large average distance from the center of gravity, a small amount of energy deposited in the first and second calorimeter layer, a broad skewness and spread of the central log ratio of the bar energies within the particle shower, and a heavy-tailed CLR of the shower energies when compared to MIP-like particles from the kurtosis. The model therefore selects energy-sparse events that do not always shower immediately after entering the calorimeter;

    \item  MIP-like events are selected using a small variance, skewness and kurtosis of the central-log ratio, a small distance to the layer centre of gravity and a comparatively large fraction of total energy deposited in the last layer, and small fractions of energy deposited in every other layer. The model therefore selects energies that are self-similar and tightly laterally distributed, with a relatively small amount of energy deposited per layer.

\end{itemize}

It is shown that the variable correlations match physical expectations: electrons are selected as longitudinally short, laterally energy-dense showers starting immediately as they enter the calorimeter;  hadrons are selected as longitudinally longer, laterally broad, energy-sparse showers expected to shower after some distance traveling through the calorimeter, and MIP-like events are very tightly distributed, reach the back of the calorimeter, and have a longitudinal center-of-gravity near the center of the calorimeter. These are all in line with physical expectations.

\begin{figure}
    \centering
    \includegraphics[width=\linewidth]{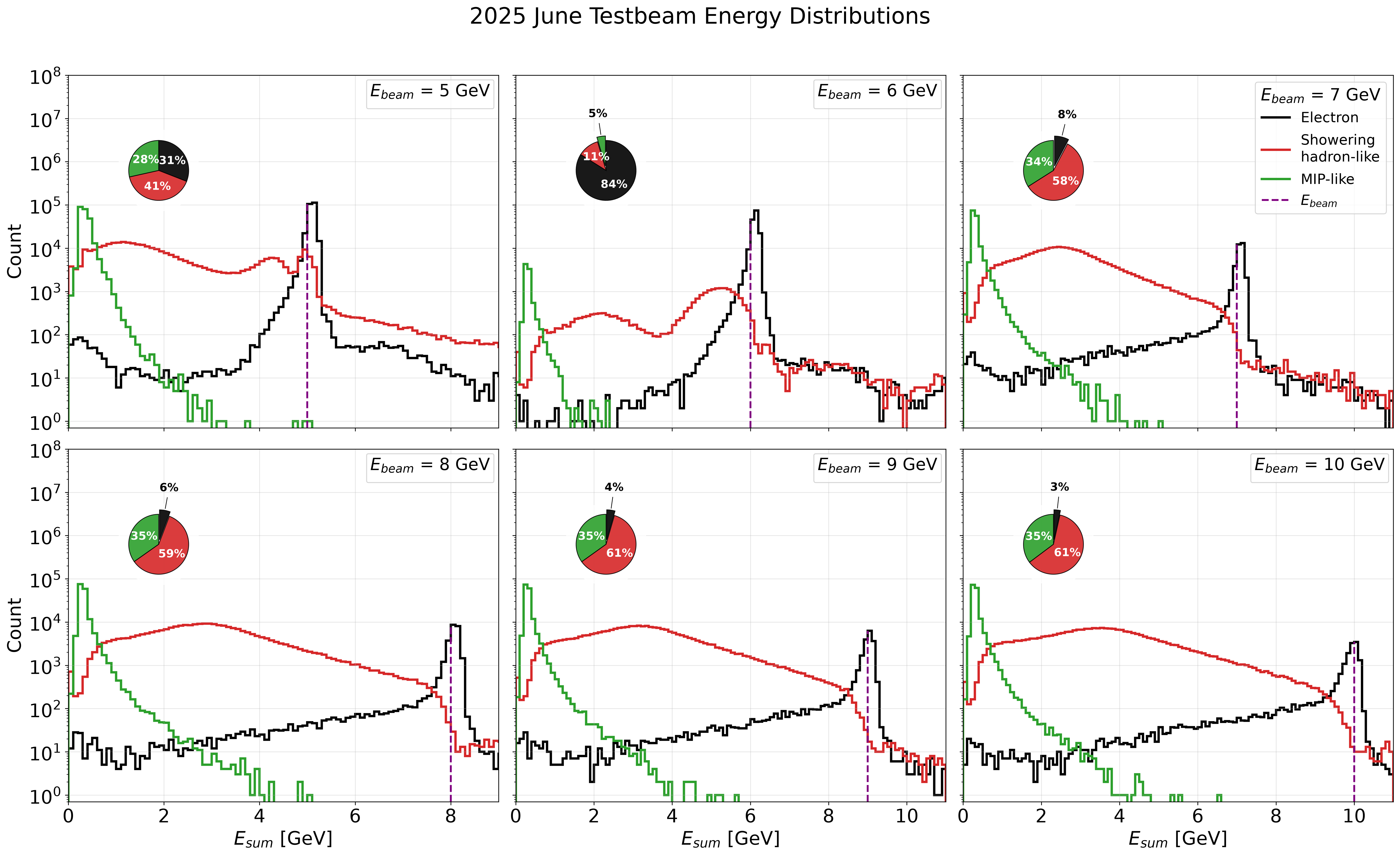}
    \caption{Energy sum distributions for the 2025 Testbeam data, separated by the PID. The \qtyrange{5}{7}{\giga \electronvolt} and \qtyrange{8}{10}{\giga \electronvolt} are shown on the top and bottom rows, respectively. The black, green, and red lines show the electron, showering hadron-like, and MIP-like events, respectively. The fraction of events in the dataset is shown by the color-coded pie chart in the top left corner. The expected electron particle energies are shown as the purple dashed lines. } 
    \label{fig:fig_energy_breakdown_grid_clr}
\end{figure}

The BDT PID was superior to the graph neural network at selecting electrons. It was therefore applied to test beam data for the \qtyrange{5}{10}{\giga \electronvolt} samples, for which contamination was most severe during the test beam. The reconstructed energy by the calorimeter is shown in Fig.~\ref{fig:fig_energy_breakdown_grid_clr}. The energy distributions of electrons are expected and match the expected skewed Gaussian core with leakage.  The sharp MIP peak shown in green indicates MIP-like events deposit roughly the same amount of energy per event. The broad continuum of energies measured for the red showering hadron-like events can be understood as a result of the partial containment and random shower start depth, which smears out the energy. The pie charts show the fraction of electrons. At beam energies greater than \qty{6}{\giga \electronvolt}, the fraction of electrons is only \qtyrange{3}{6}{\percent} and decreases with energy, illustrating the importance of PID for the analysis of this experimental data.

In summary, the supervised PID method has learned a highly effective event discrimination technique that aligns with physical expectations of interactions observed by the calorimeter.

\section{Conclusion}

Two machine learning methods were developed for PID in a highly granular crystal ECAL prototype to assess the performance of the detector in classifying basic types of events. 

Firstly, one unsupervised, contrastive method was developed for separating MIP-like punch-through $\pi^{-}$ events from showering pions using a graph neural network model with triplet loss. The classification capacity was found to be nearly perfect when classifying $\mu^{+}$ from $e^{-}$, and when applied to $\mu^{+}$ from $\pi^{-}$, it was demonstrated to select events that were in good agreement with the expectations of a MIP-like particle and showering hadron-like events. A SHAP study was used to summarize that the learned cut could be approximated using the expectation value of the log-likelihood of a Landau-Gaussian distribution over the hits in the calorimeter. This method therefore allowed the development of a bias-free cut that could be readily interpreted. 

Secondly, a supervised classifier using a graph neural network and a boosted decision tree was implemented to classify electrons, showering hadrons, and MIP-like events. Both classifiers achieved high accuracy using independent methods. Misclassification of showering hadrons was the most common of all the classes. The performance in both cases was comparable. The boosted decision tree was studied using the SHAP method and found to agree with expectations on the behavior of each event type. The BDT was applied to data and recovered the expected distributions of energy of each event type.

In summary, this study illustrates the potential for future calorimeters to achieve excellent PID performance even when the granularity is relatively coarse. It motivates further research into broader PID, such as whether or not a wider range of particles (e.g., photons, kaons, pions, protons/neutrons) can be identified by their energy deposits in a highly granular HCAL + ECAL system. It is noted as a caveat that photons were not included in this study, as its primary purpose is to purify experimental test-beam data, in which photons are unlikely to be observed. Furthermore, the effects of PID on Particle Flow reconstruction could be explored, with the goal of improving track-cluster association and therefore jet energy resolution in future Higgs factory experiments. Additionally, the unsupervised method presented can be applied beyond calorimetry to the discrimination of signal and background.

\section*{Acknowledgements}
We gratefully acknowledge the CERN management for its support and hospitality and the accelerator staff for reliable and efficient operation of the test beam. 
This research was supported by the National Key Research and Development Program of China (Grant Number 2023YFA1606300) and  MOST National Key Programme for S\&T Research and Development (Grant Number 2023YFA1606303).

\clearpage

\begin{appendices}

\section{Non-Standard Probability Density Functions}\label{sec:non_standard_distributions}

The double-sided Crystal Ball function (DSCB) is a piecewise distribution described by Eq.~\ref{eq:DSCB}:

\begin{subequations}
    \begin{equation}
    \operatorname{DSCB}(x;\mu,\sigma,\alpha_{L},n_{L},\alpha_{R},n_{R})=\begin{cases}A_{L}\cdot(B_{L}-\frac{x-\mu}{\sigma})^{-n_{L}},&  \frac{x-\mu}{\sigma}<-\alpha_{L}\\\exp(-\frac{(x-\mu)^{2}}{2\sigma^{2}}),&-\alpha_{L}\leqslant  \frac{x-\mu}{\sigma}\leqslant\alpha_{R}\\A_{R}\cdot(B_{R}-\frac{x-\mu}{\sigma})^{-n_{R}},& \frac{x-\mu}{\sigma}>\alpha_{R}\end{cases}.
    \label{eq:DSCB}
    \end{equation}

    \begin{equation}
    A_{L / R}=\left( \frac{n_{L / R}} {\left| \alpha_{L / R} \right|} \right)_{L / R}^{n_{L/ R}} \cdot\operatorname{e x p} \left(-\frac{\left| \alpha_{L / R} \right|^{2}} {2} \right)
    \end{equation}

    \begin{equation}
        B_{L / R}=\frac{n_{L / R}} {\left| \alpha_{L / R} \right|}-\left| \alpha_{L / R} \right|
    \end{equation}
\end{subequations}

where $\mu$ and $\sigma$ represent a normally-distributed core region bounded by $\alpha_{L/R}$ and with power law tails defined by $n_{L/R}$. This distribution is used to fit unknown distributions with a core fraction that is normally distributed with heavy or asymmetrical tails. It is used in this study to fit heavy-tailed distributions with a predominantly Gaussian core without requiring selection cuts, and to estimate the valid normal regions of those distributions.

\clearpage

\section{Supplementary Figures and Tables}

\begin{table}[htbp]
\caption{Counts of unique simulated events by particle, event class, and beam energy in simulation.}
\label{tab:event_counts_sim}

\captionsetup[subtable]{labelformat=simple, labelsep=none}  
\renewcommand{\thesubtable}{}                               

\begin{subtable}{\textwidth}
\centering
\caption{\qtyrange{0.1}{0.8}{\giga \electronvolt}}
\begin{tabular}{llrrrrrrrr}
\toprule
$E_{\mathrm{beam}}$ [GeV] &  & 0.1 & 0.2 & 0.3 & 0.4 & 0.5 & 0.6 & 0.7 & 0.8 \\
Particle & Event Class & & & & & & & &  \\
\midrule
$\mu^{+}$ & MIP-like & 62193 & 107287 & 118085 & 123768 & 125756 & 127980 & 129136 & 130083 \\
$e^{-}$ & Electron & 27177 & 36169 & 39328 & 40552 & 41760 & 42332 & 42786 & 43090 \\
\multirow[t]{2}{*}{$\pi^{-}$} & MIP-like & 5676 & 8227 & 12871 & 19214 & 20674 & 21129 & 21282 & 20397 \\
 & Shr. Hadron & 9955 & 24769 & 24212 & 20186 & 20329 & 20346 & 21087 & 21777 \\
\bottomrule
\end{tabular}
\end{subtable}

\vspace{1em}

\begin{subtable}{\textwidth}
\centering
\caption{\qtyrange{0.9}{4}{\giga \electronvolt}}
\begin{tabular}{llrrrrrrrr}
\toprule
$E_{\mathrm{beam}}$ [GeV] &  & 0.9 & 1 & 1.5 & 2 & 2.5 & 3 & 3.5 & 4 \\
Particle & Event Class & & & & & & & &  \\
\midrule
$\mu^{+}$ & MIP-like & 130215 & 131409 & 132617 & 133560 & 134033 & 134449 & 134781 & 134662 \\
$e^{-}$ & Electron & 43437 & 43682 & 44066 & 44626 & 44458 & 44765 & 44665 & 44971 \\
\multirow[t]{2}{*}{$\pi^{-}$} & MIP-like & 20224 & 19633 & 19878 & 19429 & 18735 & 18174 & 17470 & 16850 \\
 & Shr. Hadron & 22482 & 23007 & 23568 & 24161 & 25047 & 25924 & 26929 & 27326 \\
\bottomrule
\end{tabular}
\end{subtable}

\vspace{1em}

\begin{subtable}{\textwidth}
\centering
\caption{\qtyrange{4.5}{8}{\giga \electronvolt}}
\begin{tabular}{llrrrrrrrr}
\toprule
$E_{\mathrm{beam}}$ [GeV] &  & 4.5 & 5 & 5.5 & 6 & 6.5 & 7 & 7.5 & 8 \\
Particle & Event Class & & & & & & & &  \\
\midrule
$\mu^{+}$ & MIP-like & 134683 & 134628 & 135395 & 135345 & 135869 & 135421 & 135527 & 135240 \\
$e^{-}$ & Electron & 45067 & 44920 & 45037 & 44939 & 44994 & 44838 & 44970 & 44943 \\
\multirow[t]{2}{*}{$\pi^{-}$} & MIP-like & 16384 & 16100 & 16163 & 15691 & 15896 & 15462 & 15438 & 15732 \\
 & Shr. Hadron & 27872 & 28073 & 28273 & 28652 & 28455 & 28805 & 28917 & 28593 \\
\bottomrule
\end{tabular}
\end{subtable}

\vspace{1em}

\begin{subtable}{\textwidth}
\centering
\caption{\qtyrange{8.5}{10}{\giga \electronvolt}}
\begin{tabular}{llrrrr}
\toprule
$E_{\mathrm{beam}}$ [GeV] & & 8.5 & 9 & 9.5 & 10 \\
Particle & Event Class & & & &  \\
\midrule
$\mu^{+}$ & MIP-like & 135841 & 135378 & 135819 & 135382 \\
$e^{-}$ & Electron & 45033 & 45266 & 44757 & 44986 \\
\multirow[t]{2}{*}{$\pi^{-}$} & MIP-like & 15773 & 15841 & 15632 & 15890 \\
 & Shr. Hadron & 28715 & 29019 & 29228 & 28853 \\
\bottomrule
\end{tabular}
\end{subtable}

\end{table}

\begin{table}[htbp]
\caption{Event-class breakdown of the 2025 June Testbeam data in the range of electron energies from \qtyrange{5}{10}{\giga \electronvolt}}
\label{tab:event_counts_data}
\centering
\begin{tabular}{lrrrrrr}
\toprule
$E_{\mathrm{beam}}$ [GeV] & 5 & 6 & 7 & 8 & 9 & 10 \\
\midrule
Electron & 270,092 & 161,455 & 36,537 & 27,282 & 21,048 & 16,334 \\
Shr.\ Hadron & 354,330 & 22,120 & 275,401 & 280,530 & 285,193 & 287,792 \\
MIP-like & 248,677 & 8,873 & 160,673 & 164,181 & 164,086 & 165,511 \\
\bottomrule
\end{tabular}
\end{table}

\begin{figure}
\centering
\includegraphics[width=0.9\linewidth]{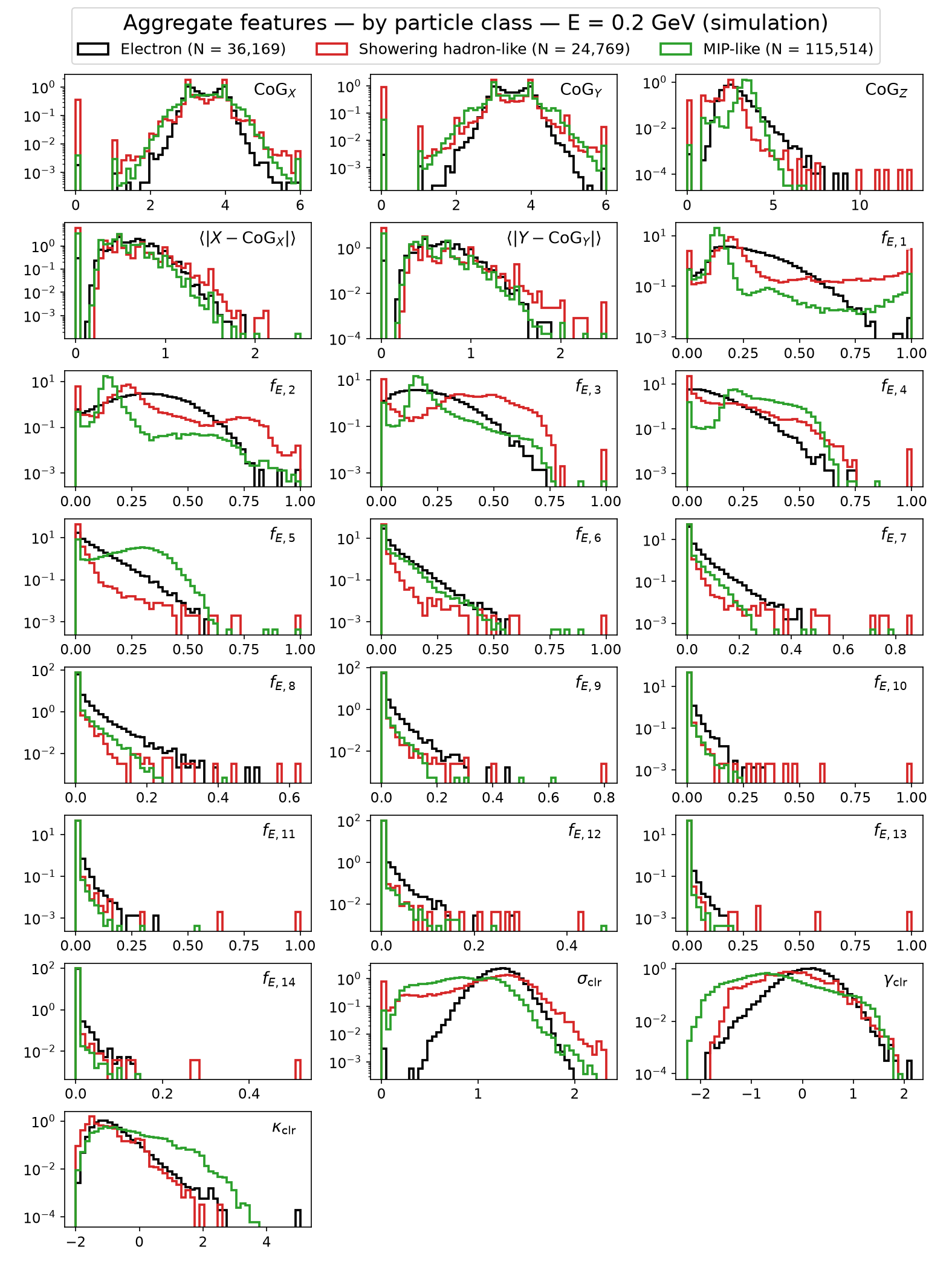}
\caption{Distributions of event-level variables used in the BDT classifier for the  \qty{0.2}{\giga \electronvolt} simulation sample. The black, red, and green distributions represent electron, MIP-like, and showering hadron-like events. The symbols shown correspond to Table~\ref{tab:bdt_params}.}
\label{fig:fig_BDT_features_by_class_E0p2}
\end{figure}

\begin{figure}
\centering
\includegraphics[width=0.9\linewidth]{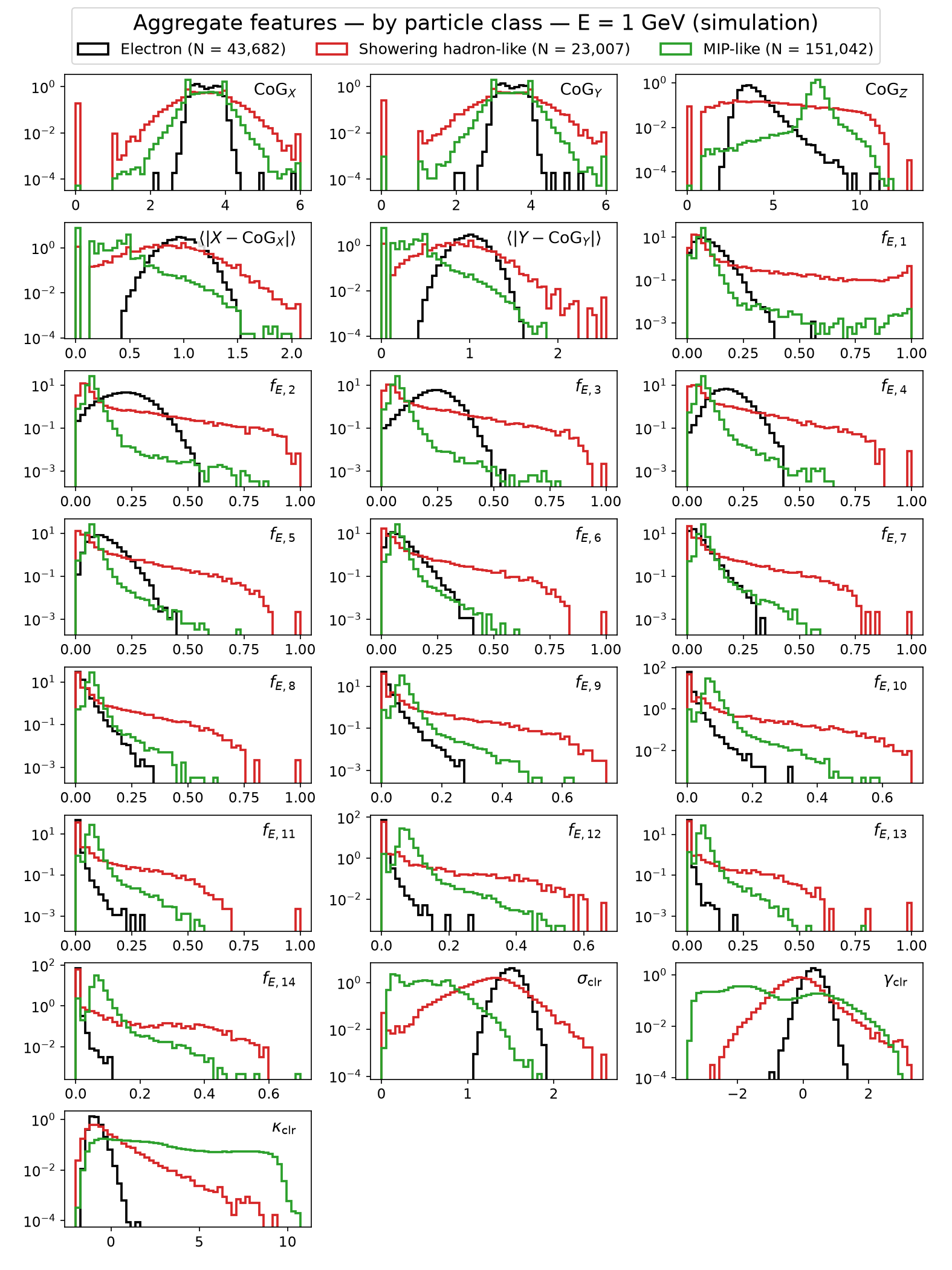}
\caption{Distributions of event-level variables used in the BDT classifier for the  \qty{1}{\giga \electronvolt} simulation sample. Else, as in Fig.~\ref{fig:fig_BDT_features_by_class_E0p2}.}
\label{fig:fig_BDT_features_by_class_E1}
\end{figure}

\begin{figure}
\centering
\includegraphics[width=0.9\linewidth]{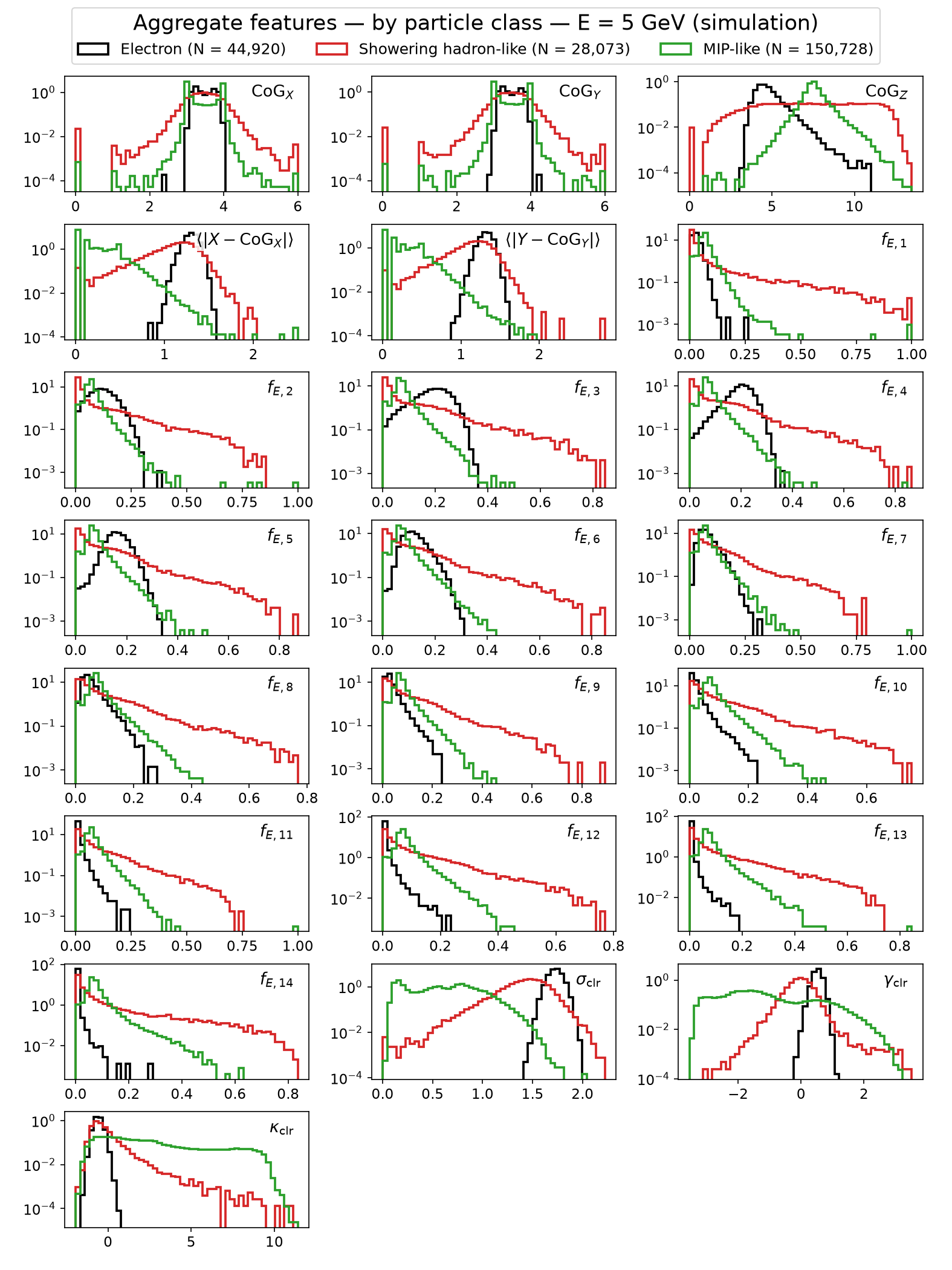}
\caption{Distributions of event-level variables used in the BDT classifier for the  \qty{5}{\giga \electronvolt} simulation sample. Else, as in Fig.~\ref{fig:fig_BDT_features_by_class_E0p2}.}
\label{fig:fig_BDT_features_by_class_E5}
\end{figure}

\begin{figure}
\centering

\begin{subfigure}{0.45\linewidth}
\centering
\includegraphics[width=\linewidth]{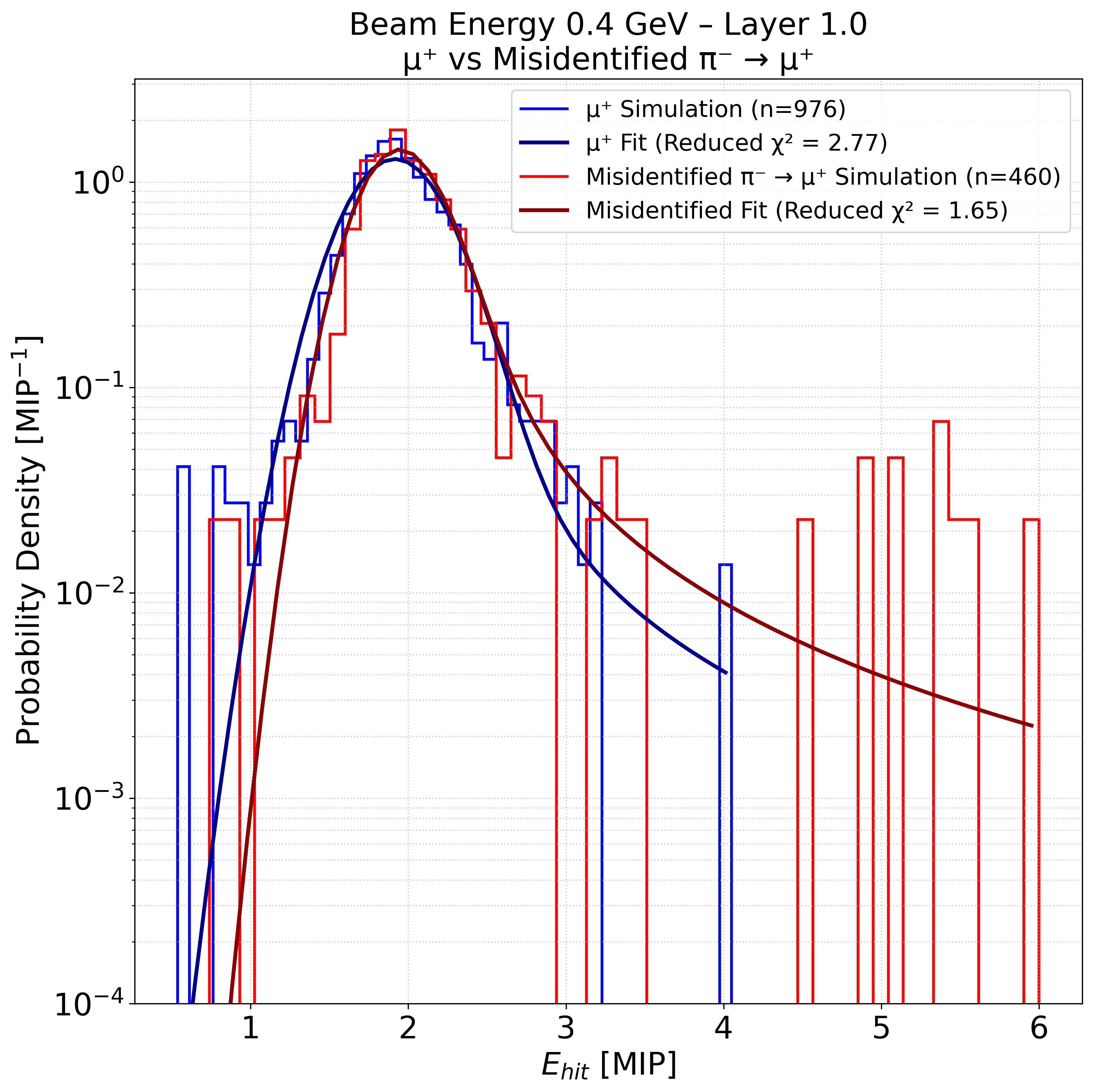}
\caption{}
\label{fig:fig_landau_fit_0p4GeV_Layer1_mu+_vs_pi-.png}
\end{subfigure}
\hfill
\begin{subfigure}{0.45\linewidth}
\centering
\includegraphics[width=\linewidth]{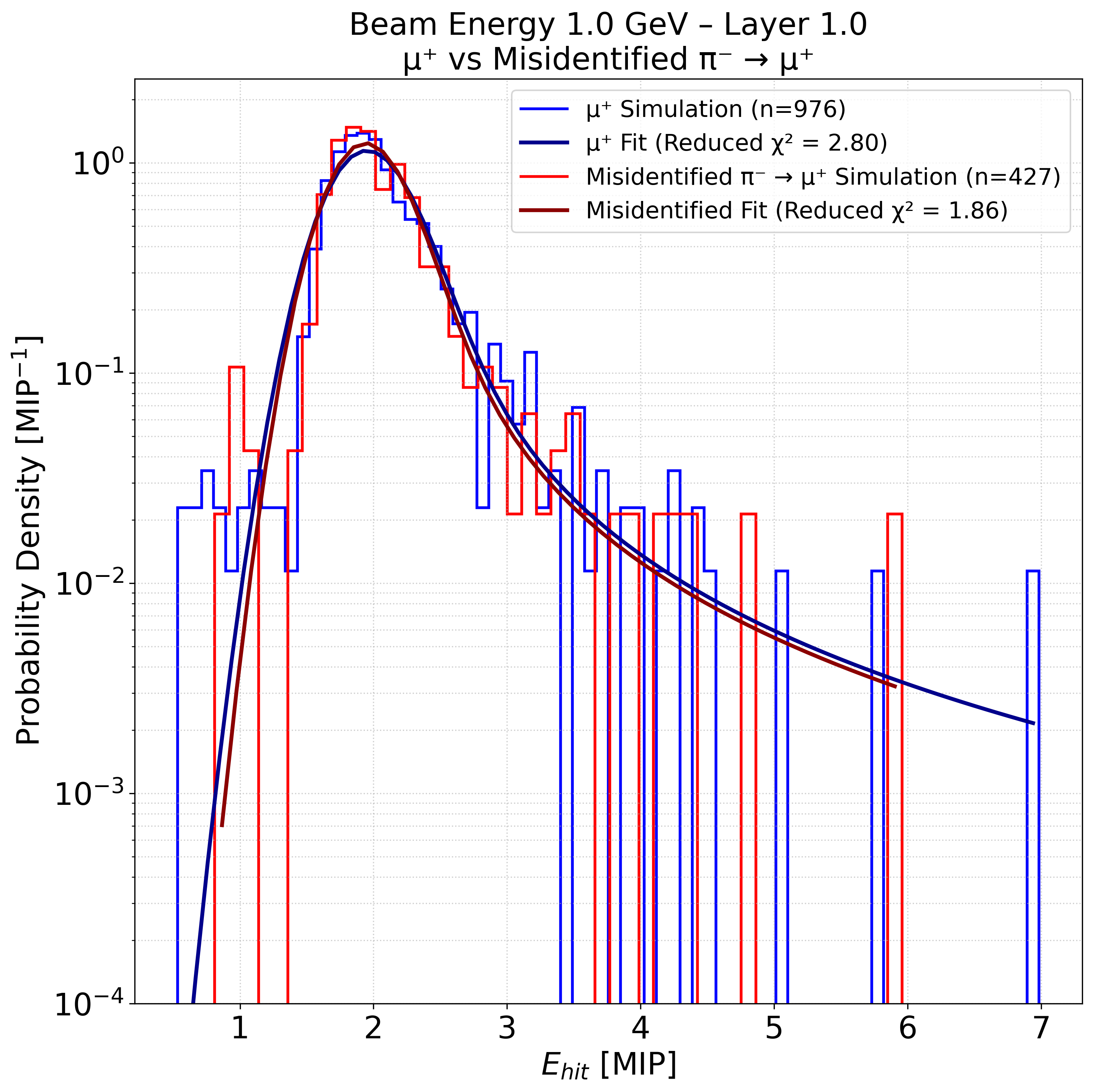}
\caption{}
\label{fig:fig_landau_fit_1p0GeV_Layer1_mu+_vs_pi-.png}
\end{subfigure}

\vfill
\begin{subfigure}{0.45\linewidth}
\centering
\includegraphics[width=\linewidth]{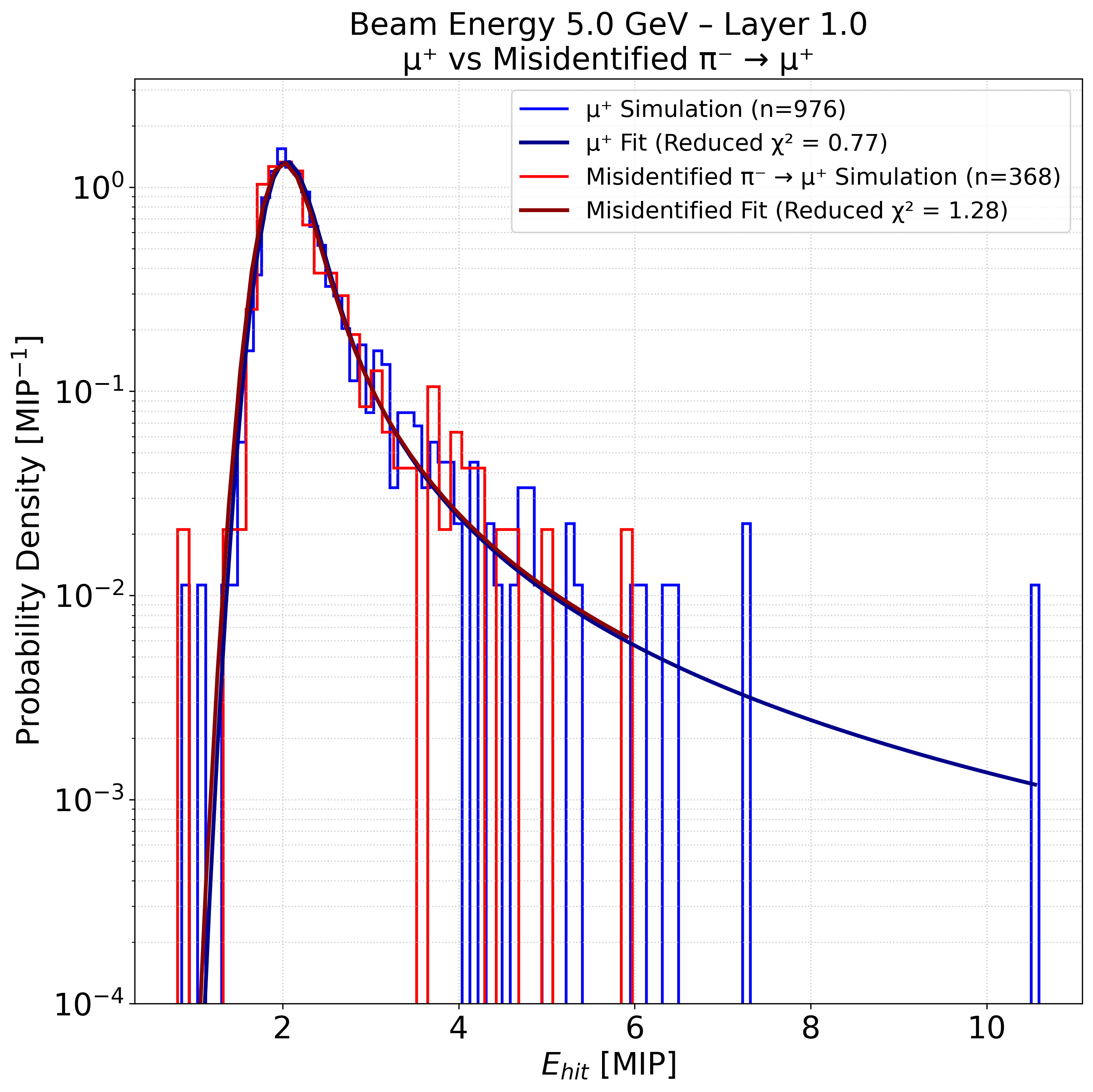}
\caption{}
\label{fig:fig_landau_fit_5p0GeV_Layer1_mu+_vs_pi-.png}
\end{subfigure}
\hfill
\begin{subfigure}{0.45\linewidth}
\centering
\includegraphics[width=\linewidth]{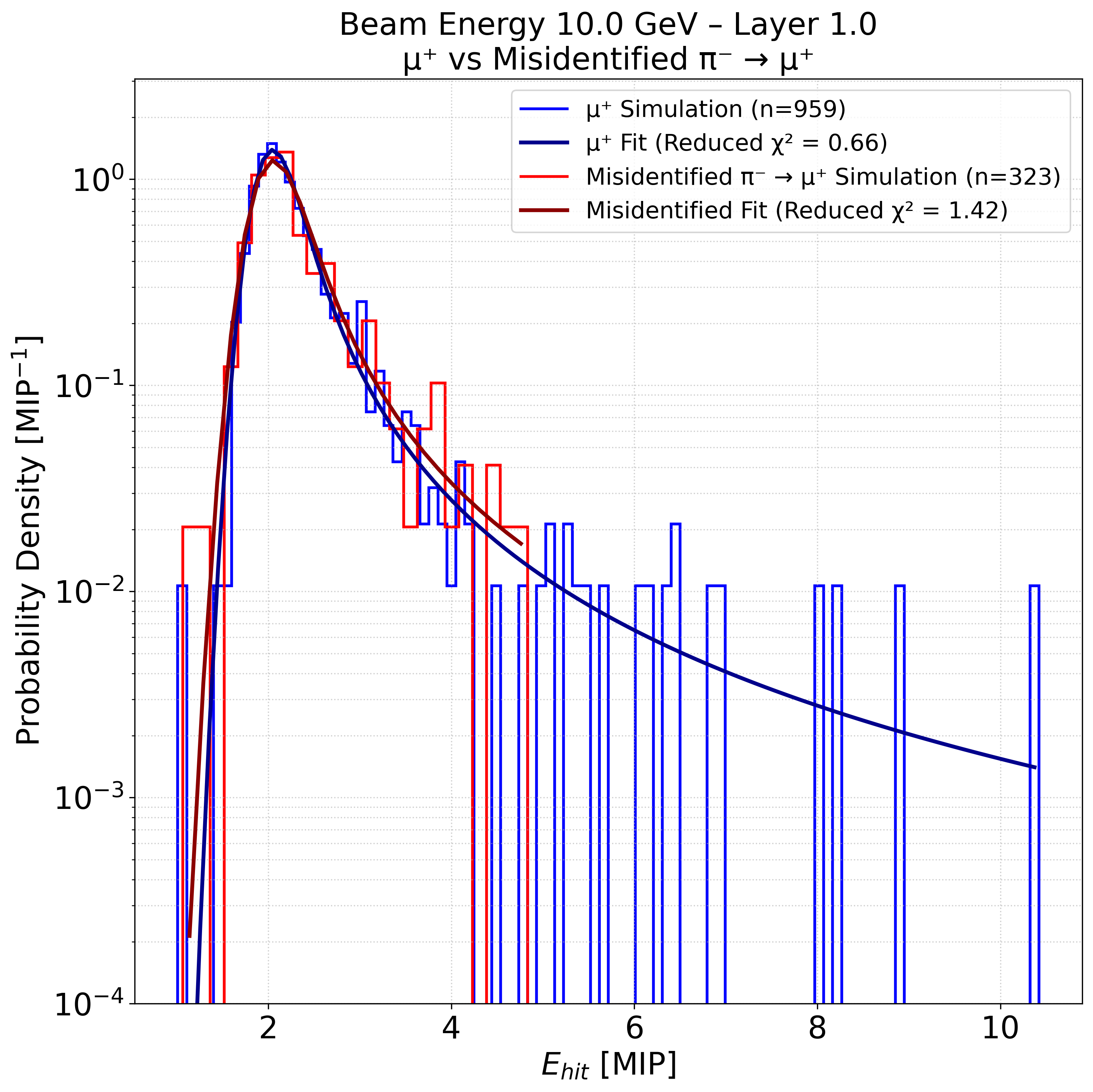}
\caption{}
\label{fig:fig_landau_fit_10.0GeV_Layer1_mu+_vs_pi}
\end{subfigure}

\caption{Example fits of the Landau-Gaussian distribution to the hit energy distribution in the first layer of the calorimeter. The light blue and red lines indicate the distributions of the total energy deposited in a bar in the first calorimeter layer, for $\mu^{+}$ and misidentified $\pi^{-}$, respectively. The dark blue and dark red lines indicate fits of the Landau-Gaussian distribution from Eq.~\ref{eq:landau_gaus}, with the same color coding. Figs.~\ref{fig:fig_landau_fit_0p4GeV_Layer1_mu+_vs_pi-.png}-
~\ref{fig:fig_landau_fit_1p0GeV_Layer1_mu+_vs_pi-.png} show the \qty{0.4}{\giga \electronvolt},+ \qty{1}{\giga \electronvolt}, \qty{5}{\giga \electronvolt} and \qty{10}{\giga \electronvolt} datasets. }
\label{fig:fig_landau_fit}

\end{figure}

\begin{figure}
\centering
\includegraphics[width=0.6\linewidth]{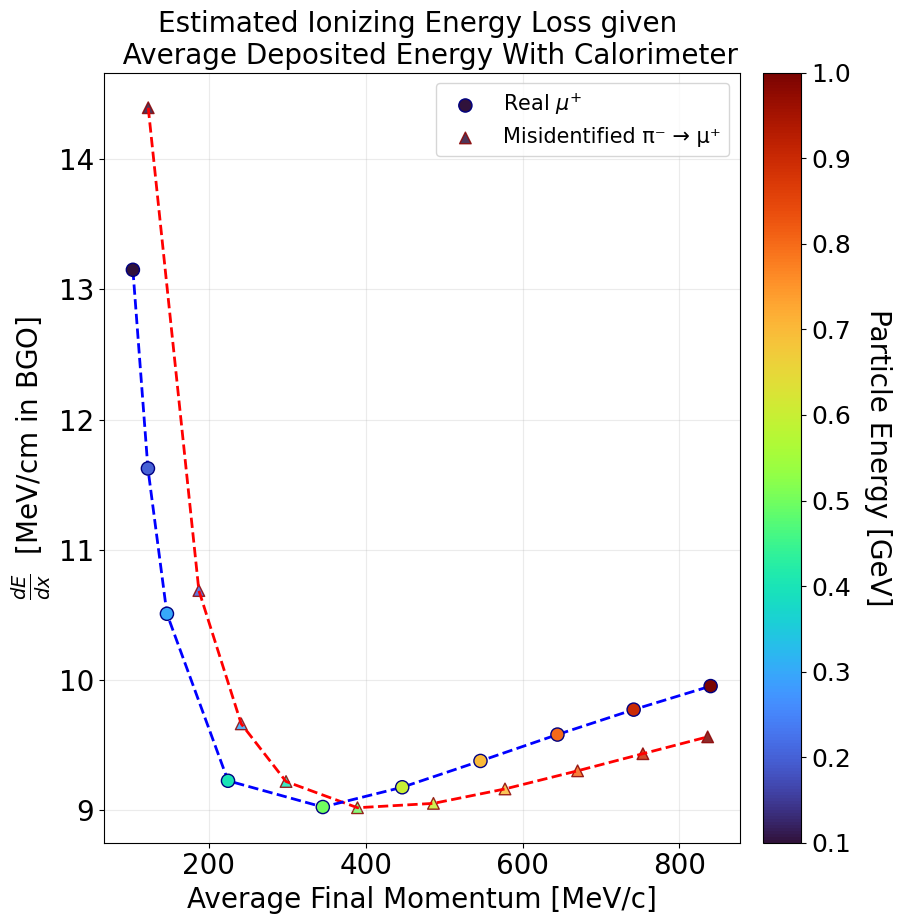}
\caption{Estimated ionizing energy loss in BGO as a function of average final momentum, estimated from average total reconstructed energy, for $  \mu^+  $ and misidentified $ \pi^- $ particles, indicated by blue dashed lines and circle markers, and the misclassified $\pi^{-}$ as red dashed lines and triangle markers. Values for the $\frac{Z}{A}$ ratio, density, and ionization potential for BGO were taken from Ref.~\cite{bakken_ionization_1988}. The density effect was not included.}
\label{fig:fig_bethe_bloch_beta2_region}
\end{figure}

\end{appendices}

\clearpage

\printbibliography

\end{document}